\documentclass[twocolumn,prd,aps,superscriptaddress,preprintnumbers,tightenlines,showpacs,nofootinbib,eqsecnum,amsfonts,amsmath]{revtex4}
\pdfoutput=1
\usepackage{amsmath,amssymb,mathrsfs}
\usepackage{graphicx}
\usepackage[nolist,nohyperlinks]{acronym}
\usepackage[usenames,dvipsnames]{xcolor}
\usepackage{float}
\usepackage{placeins}
\usepackage[utf8]{inputenc}

\newcommand{\AEI}{\affiliation{Max Planck Institute for Gravitational Physics (Albert Einstein Institute), Am M\"uhlenberg 1, Potsdam 14476, Germany}}

\newcommand{\APC}{\affiliation{APC, AstroParticule et Cosmologie, Universit\'e Paris Diderot, CNRS/IN2P3, CEA/Irfu, Observatoire de Paris, Sorbonne Paris Cit\'e, France}}

\def\be{\begin{equation}}
\def\ee{\end{equation}}
\def\bea{\begin{eqnarray}}
\def\eea{\end{eqnarray}}
\newcommand{\bes}{\begin{subequations}}
\newcommand{\ees}{\end{subequations}}

\DeclareMathOperator{\range}{range}
\DeclareMathOperator{\rank}{rank}

\allowdisplaybreaks

\graphicspath{{./Graphics/}}

\begin{document}

\pagenumbering{arabic}

\title{Frequency domain reduced order model of aligned-spin effective-one-body waveforms with higher-order modes}

\author{Roberto Cotesta}
\email{roberto.cotesta@aei.mpg.de}
\AEI

\author{Sylvain Marsat}
\APC

\author{Michael P{\"u}rrer}
\AEI

\date{\today}

\begin{abstract}
We present a frequency domain reduced order model (ROM) for the aligned-spin effective-one-body (EOB) model for binary black holes (BBHs) \texttt{SEOBNRv4HM} that includes the spherical harmonics modes $(\ell, |m|) = (2,1),(3,3),(4,4),(5,5)$ beyond the dominant $(\ell, |m|) = (2,2)$ mode. These higher modes are crucial to accurately represent the waveform emitted from asymmetric BBHs.
We discuss a decomposition of the waveform, extending other methods in the literature, that allows us to accurately and efficiently capture the morphology of higher mode waveforms.
We show that the ROM is very accurate with median (maximum) values of the unfaithfulness against \texttt{SEOBNRv4HM} lower than $0.001\% (0.03\%)$ for total masses in $[2.8,100] M_\odot$. For a total mass of $M = 300 M_\odot$ the median (maximum) value of the unfaithfulness increases up to $0.004\% (0.17\%)$. This is still at least an order of magnitude lower than the estimated accuracy of SEOBNRv4HM compared to numerical relativity simulations.
The ROM is two orders of magnitude faster in generating a waveform compared to \texttt{SEOBNRv4HM}. Data analysis applications typically require $\mathcal{O}(10^6-10^8)$ waveform evaluations for which SEOBNRv4HM is in general too slow. The ROM is therefore crucial to allow the SEOBNRv4HM waveform to be used in searches and Bayesian parameter inference.
We present a targeted parameter estimation study that shows the improvements in measuring binary parameters when using waveforms that includes higher modes and compare against three other waveform models.
\end{abstract}

\maketitle


\acrodef{LSC}[LSC]{LIGO Scientific Collaboration}
\acrodef{BH}{black hole}
\acrodef{NS}{neutron star}
\acrodef{PN}{Post-Newtonian}
\acrodef{BBH}{binary black-hole}
\acrodef{BNS}{binary neutron-star}
\acrodef{NSBH}{neutron-star black-hole}
\acrodef{EOB}{effective-one-body}
\acrodef{NR}{numerical relativity}
\acrodef{GW}{gravitational-wave}
\acrodef{PSD}{power spectral density}
\acrodef{aLIGO}{Advanced Laser interferometer Gravitational-Wave Observatory}
\acrodef{AZDHP}{aLIGO zero detuned high power density}
\acrodef{GR}{general relativity}
\acrodef{PE}{parameter estimation}
\acrodef{LAL}{LIGO algorithm library}
\acrodef{TPI}{tensor-product interpolant}
\acrodef{SVD}{singular value decomposition}
\acrodef{SNR}{signal to noise ratio}
\acrodef{ODE}{ordinary differential equation}
\acrodef{ROM}{reduced order model}

\section{Introduction}

In the past five years \ac{GW} observations~\cite{Abbott:2016blz,TheLIGOScientific:2016pea,LIGOScientific:2018mvr} have opened up a new window to the Universe. In the first two observing runs of the advanced LIGO~\cite{TheLIGOScientific:2014jea} and Virgo~\cite{TheVirgo:2014hva} detectors ten confident detections of \acp{BBH} and one \ac{BNS} were made~\cite{LIGOScientific:2018mvr} and tens of detection candidates~\cite{gracedb-O3-superevents} have been identified so far in the third observing run of this network, among them another confident detection of a binary neutron star system~\cite{Abbott:2020uma}.
Both the detection and inference of binary parameters for these compact binaries rely heavily on our knowledge of the gravitational waveform emitted in these coalescences as encapsulated in parametrized models of \acp{GW}. The construction of stochastic template banks and the Bayesian inference of binary parameters routinely require tens to hundreds of millions of waveform evaluations~\cite{Harry:2009ea,Manca:2009xw, Veitch:2014wba,Ashton:2018jfp}. At the same time the phasing of the \acp{GW} needs to be tracked to an accuracy better than a fraction of a wave cycle to avoid missing events or mis-measuring binary parameters. Therefore, waveform models need to be fast and accurate to extract the binary properties imprinted in the emitted \acp{GW}.

Inspiral-merger-ringdown models of \acp{GW} from coalescing \ac{BH} binaries have traditionally been constructed in the \ac{EOB}~\cite{Buonanno:1998gg,Buonanno:2000ef,Damour:2000we,Damour:2001tu,Damour:2008gu,Pan:2011gk,Taracchini:2012ig,Taracchini:2013rva,Pan:2013rra,Damour:2014sva,Nagar:2015xqa,Bohe:2016gbl,Babak:2016tgq,Nagar:2018zoe} or phenomenological approaches~\cite{Ajith:2011,Santamaria:2010yb,Hannam:2013oca,Husa:2015iqa,Khan:2015jqa,Mehta:2017jpq,London:2017bcn,Khan:2018fmp,Khan:2019kot,Pratten:2020fqn}.
\ac{EOB} models incorporate physical descriptions of the inspiral, merger, and ringdown parts of \ac{BBH} coalescences. \ac{PN} solutions for the inspiral are re-summed and connected with an analytic description of the merger waveform which is tuned to data from \ac{NR} simulations~\cite{Pretorius:2005gq,Campanelli:2005dd,Baker:2005vv,
Bruegmann:2006at,Centrella:2010mx,Mroue:2013xna,Jani:2016wkt,Healy:2017psd,Boyle:2019kee}. \ac{EOB} models are posed as an initial value problem for a complicated system of \acp{ODE} describing the dynamics of a compact binary. The emitted \acp{GW} are then computed from this orbital dynamics. \ac{EOB} models provide accurate and generic descriptions of the \acp{GW}.
The integration of the \acp{ODE} requires small time steps to obtain an accurate solution and especially for the long waveforms produced by low mass compact binaries can take on the order of hours, and thus be too slow for practical data analysis applications~\footnote{
  A faster method has been proposed, restricted to systems with spins aligned with the orbital angular momentum~\cite{Nagar:2018gnk}.
}.

Surrogate or reduced order modeling techniques~\cite{Field:2013cfa,Purrer:2014fza,Purrer:2015tud,Blackman:2015pia,Blackman:2017pcm,Blackman:2017dfb,Lackey:2018zvw,Doctor:2017csx,Setyawati:2019xzw} provide established methods for accelerating slow waveforms while retaining very high accuracy. These techniques have been successfully applied to \ac{EOB}~\cite{Field:2013cfa,Purrer:2014fza,Purrer:2015tud,Bohe:2016gbl,Lackey:2018zvw} and \ac{NR}~\cite{Blackman:2015pia,Blackman:2017pcm,Blackman:2017dfb,Varma:2018mmi,Varma:2019csw} waveforms. They work by decomposing waveforms from a training set in orthonormal bases on sparse grids in time or frequency, and interpolating or fitting the resulting waveform data pieces over the binary parameter space. The result is a smooth, accurate (as tested against an independent validation set), and fast to evaluate (compared to the original waveform data) \ac{GW} model. These surrogate models have proven invaluable for \ac{GW} data analysis.

In this paper we present a \ac{ROM} for \acp{GW} from coalescing binaries with spins aligned with the orbital angular momentum which include the most important higher harmonics of the waveform in addition to the dominant $(\ell. m) = (2, \pm 2)$ spherical harmonic mode, as described by the \texttt{SEOBNRv4HM} model~\cite{Cotesta:2018fcv}.
Higher harmonics in the expansion of the gravitational waveform become important for asymmetric and massive compact binaries~\cite{Brown:2012nn,Capano:2013raa,Harry:2017weg, Varma:2014jxa,Varma:2016dnf,Kalaghatgi:2019log}. The model we construct here, \texttt{SEOBNRv4HM\_ROM}, includes the $(\ell, |m|) = (2, 1), (3, 3), (4, 4), (5, 5)$ modes. We show that \texttt{SEOBNRv4HM\_ROM} has a mismatch less than $\mathcal{O}(0.1\%)$ with \texttt{SEOBNRv4HM} and that it accelerates waveform evaluation by a factor 100 -- 200.

This paper is organized as follows. In Sec.~\ref{sec:SEOBNRv4HM} we give a brief description of the time-domain \texttt{SEOBNRv4HM} model. In Sec.~\ref{sec:techniques} we discuss various techniques we use to build the ROM, notably waveform conditioning in Sec.~\ref{sec:preparation_and_decomposition_of_wf_data}.
We continue with a summary of the basis construction and decomposition in Sec.~\ref{sec:SVD}, tensor product interpolation in Sec.~\ref{sec:TPI}. Domain decomposition in frequency and in parameter space is discussed in Sec.~\ref{sec:patching-in-frequency} and Sec.~\ref{sec:patching_parameter_space}, respectively.
We summarize how we connect the ROM with \ac{PN} solutions at low frequency in Sec.~\ref{sec:hybridization}.
We present results in Sec. ~\ref{sec:results} where we demonstrate the accuracy of the ROM in Sec~\ref{sec:accuracy}, and its computational efficiency in Sec.~\ref{sec:computational_performance}. We showcase a parameter estimation application in Sec.~\ref{sec:PEsec}.
Finally we conclude in Sec.~\ref{sec:conclusion}.

\section{The \texttt{SEOBNRv4HM} model}
\label{sec:SEOBNRv4HM}

The gravitational wave signal emitted by a coalescing binary black hole is usually divided into three different regimes: \textit{inspiral}, \textit{merger} and \textit{ringdown}. During the inspiral the two black holes move at a relative speed $v$ that is small compared to the speed of light $c$, therefore the solution to the two body problem can be found using a perturbative expansion in the small parameter $v/c$, the so-called \ac{PN} expansion~\cite{Blanchet:2013haa}. At some point, during the evolution of the binary system, the parameter $v/c$ ceases to be small and the \ac{PN} expansion is not valid anymore. This happens roughly at the innermost stable circular orbit (ISCO) and demarcates the end of the inspiral and the beginning of the merger regime. The signal in this regime can only be computed using \ac{NR} simulations that solve Einstein's equations for a \ac{BBH} system, fully numerically. Finally, in the ringdown regime, the perturbed black hole formed after the merger of the binary emits gravitational waves at frequencies that can be computed within the black hole perturbation theory formalism~\cite{Berti:2009kk}.

The EOB formalism, first introduced by Buonanno and Damour in Refs.~\cite{Buonanno:1998gg,Buonanno:2000ef}, provides a natural framework to combine these three regimes and produce a complete waveform with inspiral, merger and ringdown. Within the EOB formalism the \ac{PN} conservative dynamics of a \ac{BBH} system during the inspiral is resummed in terms of the dynamics of a test particle with an effective mass and spin around a deformed Kerr metric. This improved conservative dynamics is combined with a resummed energy flux~\cite{Pan:2010hz,PhysRevD.79.064004,PhysRevD.76.064028} to produce an inspiral waveform that is close to \ac{NR} solutions. To improve the agreement with \ac{NR} waveforms the \ac{EOB} conservative dynamics is also \textit{calibrated} using information from \ac{NR} simulations~\cite{Taracchini:2012ig,Taracchini:2013rva}. In the EOB waveform the merger and ringdown part is built using a phenomenological fit produced using informations from \ac{NR} waveforms and black hole perturbation theory~\cite{Nagar:2016iwa,Bohe:2016gbl}. NR-tuned versions of EOB models are usually referred to as EOBNR.

In this paper we focus on the \texttt{SEOBNRv4HM}~\cite{Cotesta:2018fcv} model, an extension of \texttt{SEOBNRv4}~\cite{Bohe:2016gbl} that includes the modes $(\ell, |m|) = (2, 1), (3, 3), (4, 4), (5, 5)$ in the waveform  in addition to the $(\ell, |m|) = (2, 2)$ mode already present in  \texttt{SEOBNRv4}. This model assumes spins aligned or anti-aligned with the direction perpendicular to the orbital plane $\hat L_N$, and we define the dimensionless spin parameter for \ac{BH} $i$ as $\chi_i = \vec S_i \cdot \hat L_N / m_i^2$, where $\vec S_i$ are the BH’s spins and $m_i$ their masses. \texttt{SEOBNRv4HM} has been validated against several NR waveforms in the mass-ratio - aligned-spin parameter space in the region $q \equiv m_1/m_2 \in [1,10]$, $\chi_{1,2} \in [-1,1]$ yielding typical mismatches of $\mathcal{O}(\leq1\%)$ for total masses in the range $[20,200] M_\odot$.

\section{Techniques for building the ROM}
\label{sec:techniques}

In this Section we describe the construction of our ROM, from the preparation of the waveforms to the reduced basis and interpolation techniques. We use techniques developed for previous ROM models~\cite{Purrer:2014fza,Purrer:2015tud,Bohe:2016gbl}, which we generalized to the higher-harmonics case. 

We start with a general discussion of how to prepare and decompose waveform data for higher mode waveforms in Sec.~\ref{sec:preparation_and_decomposition_of_wf_data}. In particular, we discuss time domain conditioning in Sec.~\ref{sec:TD-conditioning}, stationary phase approximation in Sec.~\ref{sec:SPA}, the orbital carrier phase in Sec.~\ref{sec:carrier}, the introduction of coorbital modes in Sec.~\ref{sec:coorbital_modes_and_waveform_building_blocks}, scaling of frequencies in Sec.~\ref{sec:scaling_of_frequencies}.
We summarize basis construction in Sec.~\ref{sec:SVD} and tensor product interpolation in Sec.~\ref{sec:TPI}.
We also explain how we decompose the model in both frequency range patches (see Sec.~\ref{sec:patching-in-frequency}) and parameter space patches (see Sec.~\ref{sec:patching_parameter_space}).
Hybridization with inspiral waveforms is discussed in Sec.~\ref{sec:hybridization}.

\subsection{Preparation and decomposition of waveform data}
\label{sec:preparation_and_decomposition_of_wf_data}

The waveform polarizations $h_{+}$, $h_{\times}$ are decomposed in spin-weighted spherical harmonics as
\be
	h_{+} - i h_{\times} = \sum_{\ell \geq 2} \sum_{m=-\ell}^{\ell} {}_{-2}Y_{\ell,m} h_{\ell m} \,.
\ee
The $h_{\ell m}$ are called the harmonics or simply the modes of the gravitational wave, with $h_{22}$ and $h_{2,-2}$ the dominant harmonics corresponding to quadrupolar radiation. These modes $h_{\ell m}$ are affected by convention choices: first, by the choice of polarization vectors defining $h_{+}$, $h_{\times}$, and secondly by the definition chosen for the source frame in which the waveform is described. For non-precessing systems, the $z$-axis of the source frame is taken to be the normal to the orbital plane, with a residual freedom in choosing the origin of phase. One can take two points of view for the definition of phase: either fix the definition of the source frame (for instance, imposing that the initial separation vector is along $x$) and call ``phase'' the azimuthal angle of the observer in the source frame, or fix the direction to the observer (for instance in the plane $(x,z)$) and call ``phase'' the binary's orbital phase at a given time. We can also consider the definition of the origin of time as part of the source frame definition.

During the inspiral, the individual harmonics obey a simple overall scaling with the orbital phase as
\be\label{eq:deforbitalphasing}
	h_{\ell m} \propto \exp\left[ - i m \phi_{\rm orb}\right] \,,
\ee
but this scaling does not apply post-merger where the modes are driven by their respective ringdown frequencies.

There are several challenges regarding the conditioning of higher-harmonics waveforms for the purpose of reduced order modelling. We recall that one relies on two kinds of interpolation here: one is the interpolation of waveform pieces along the tracking parameter, either time or frequency, used to compress data; the other is the interpolation across the parameter space (masses and spins) used in the internals of the ROM, either of waveform quantities directly (as in~\cite{Field:2013cfa,Blackman:2015pia,Blackman:2017pcm,Blackman:2017dfb,Lackey:2018zvw,Varma:2018mmi,Varma:2019csw} in the empirical interpolation formalism), or as in our case, of reduced basis projection coefficients~\cite{Purrer:2014fza,Purrer:2015tud,Bohe:2016gbl}. Both these interpolations require smoothness, and discrete jumps can cause significant (and non-local) errors.

As a result, zero-crossings in the subdominant harmonics $h_{\ell m}$ (as noticed in Refs.~\cite{Cotesta:2018fcv,Nagar:2020pcj}) cause difficulties for the usual amplitude/phase representation: if the envelope of a mode crosses zero with a positively defined amplitude, the phase jumps by $\pi$, a discontinuity that will harm the interpolation performed when reconstructing the waveform. Among other advantages, this is alleviated by the procedure used in~\cite{Blackman:2017pcm, Varma:2019csw} of modelling the waveform in a coorbital frame where the dominant phasing of Eq.~\eqref{eq:deforbitalphasing} has been scaled out, so that a more robust real/imaginary representation can be chosen instead; here we will use the same kind of coorbital quantities, but built in the Fourier domain.

The natural $2\pi$ degeneracy in phase also requires care when interpolating across parameter space. Discrete $2\pi$ phase jumps leave the waveforms themselves invariant, but can break the interpolation in-between waveforms. This issue is particularly relevant when dealing with numerical Fourier transforms of time-domain waveforms: when phase-unwrapping the output of the Discrete Fourier Transform starting from $f=0$, numerical noise causes the $2\pi$ interval to be essentially random. In~\cite{Purrer:2014fza,Purrer:2015tud,Bohe:2016gbl} a linear fit of the Fourier-domain phase was removed. Here we will keep time and phase alignment information throughout the conditioning procedure, so instead we will impose a given $2\pi$ range for the phase at a reference point, corresponding to the time of alignment.

Other difficulties are caused by the relative alignment of the different harmonics. Dividing the phase of the dominant $h_{22}$ mode by 2, whether in time or frequency domain, comes with an ambiguity of $\pi$ then propagated as $m \pi$ to the other modes. Such an ambiguity is not necessarily a problem if the phase alignment is done as a last step when generating the waveform (as is the case in the \texttt{IMRPhenomHM} model~\cite{London:2017bcn}); giving up the geometric interpretation of the source-frame definition, it is sufficient that a $[0,2\pi]$ range in the ``phase'' input by the user corresponds to a $[0,2\pi]$ range in geometric phase. It becomes a problem, however, when we need to interpolate across parameter space to build a ROM. In particular, when working from the Fourier domain waveform alone, we do not have access to the orbital phase (as read from trajectories) to lift these kind of degeneracies. Here we will make sure that the alignment is performed in the time domain before taking Fourier transforms, and we will further introduce an artificial carrier signal to have access to a proxy of the orbital phase in the Fourier domain.

We detail below our conditioning procedure, chosen to circumvent these issues.

\subsubsection{Time-domain conditioning}
\label{sec:TD-conditioning}

In building this ROM, we will carry along time and phase alignment information all the way to the final Fourier-domain waveforms. This is in contrast to previous Fourier-domain waveform models (\texttt{SEOBNRv4\_ROM}, \texttt{IMRPhenomD}) where the time and phase are adjusted after generating the waveform, as will be described below.

Individual harmonics are decomposed as an amplitude\footnote{In general, it would be preferable to consider $a_{\ell m}$ as a slowly-varying envelope rather than a positive amplitude, in particular allowing it to change sign, as we expect zero-crossings of certain subdominant modes like $h_{21}$.} and phase, following
\be\label{eq:hlmtd}
	h_{\ell m}(t) = a_{\ell m}(t) \exp \left[ -i \phi_{\ell m}(t)\right] \,,
\ee
with the scaling
\be\label{eq:philmscaling}
	\phi_{\ell m} = m \phi_{\rm orb} + \Delta\phi_{\ell m} \,,
\ee
In the early inspiral regime, for low frequencies, the phases $\Delta\phi_{\ell m}$ are approximately constant. We choose the same polarization convention as in~\cite{lrr-2006-4}, for which we have
\begin{align}
  \label{eq:delta_phi_constants}
	\Delta\phi_{22} &\rightarrow 0\,, \nonumber\\
	\Delta\phi_{21} &\rightarrow \frac{\pi}{2}\,, \nonumber\\
	\Delta\phi_{33} &\rightarrow -\frac{\pi}{2}\,, \nonumber\\
	\Delta\phi_{44} &\rightarrow \pi\,, \nonumber\\
	\Delta\phi_{55} &\rightarrow \frac{\pi}{2} \,,
\end{align}
in the low-frequency limit. When getting closer to merger, deviations from Eq.~\eqref{eq:delta_phi_constants} become more important. In the notations of the EOB factorized waveforms~\cite{Damour:2008gu,Pan:2011gk}, these phase deviations come from the phases $e^{i \delta_{\ell m}}$ and tail factors $T_{\ell m}$ (see Eqs.~(14) and~(21) in~\cite{Pan:2011gk}), and from non-quasicircular corrections (NQCs) close to merger (see Eq. (22) in~\cite{Pan:2011gk}).

 We choose the source frame convention for our model by imposing that its direction $x$ is along the separation vector between the two black holes $n(t_{\rm align})$, with an arbitrary time of alignment in the late inspiral $t_{\rm align} = -1000M$ (with $t = 0$ being defined as the amplitude peak of $h_{22}$). In practice, rather than using $n(t_{\rm align})$ we simply impose
\be
	\phi_{22} (t_{\rm{align}}) = 0 \,,
\ee
and we use the orbital phase $\phi_{\rm orb}$ as read from the EOB dynamics to resolve the $\pi$-ambiguity and impose $\phi_{\rm orb} \simeq 0$ rather than $\phi_{\rm orb} \simeq \pi$. These alignment properties will be reproduced, up to small numerical errors, by the reconstructed ROM waveforms.

\subsubsection{Stationary phase approximation}
\label{sec:SPA}

As we will use it to guide our conditioning procedure, we recall here the Stationary phase approximation (SPA) for waveforms with higher harmonics. First, we introduce the Fourier transform for a time-domain signal $h$ as
\be
	\tilde{h}(f) = \int dt \, e^{2i \pi f t} h(t) \,.
\ee
Note the sign difference in the exponential with respect to the more usual definition (used in particular in LAL~\cite{LALSuiteGit}). This choice is made for convenience, as it will ensure that Fourier-domain modes $\tilde{h}_{\ell m}$ with $m>0$ and $m<0$ have support at positive and negative frequencies respectively. One can come back to the LAL Fourier convention by the simple map $f \leftrightarrow -f$, which for real signals $h(t) \in \mathbb{R}$ translates as $\tilde{h}_{\rm LAL}(f) = \tilde{h}^{*}(f)$.

Let us first consider a generic signal with an amplitude and phase as $h(t) = a(t)e^{-i \phi(t)}$ and define $\omega \equiv \dot{\phi}$. The SPA is applicable under the assumptions $\left| \dot{a} / (a\omega)\right| \ll 1$, $\left| \dot{\omega} / \omega^{2}\right| \ll 1$ and $\left| (\dot{a} / a)^{2} / \dot{\omega}\right| \ll 1$, that are well verified in the inspiral. Defining a time-to-frequency correspondence $t(f)$ implicitly by
\be\label{eq:deftfSPA}
	\omega(t(f)) = 2\pi f \,,
\ee
the Fourier tranform of the signal is then $\tilde{h}_{\rm SPA}(f) = A_{\rm SPA}(f) e^{-i \Psi_{\rm SPA}(f)}$ with
\bes
\begin{align}
	A_{\rm SPA}(f) &= a(t(f)) \sqrt{\frac{2 \pi}{\dot{\omega}(t(f))}} \,, \\
	\Psi_{\rm SPA}(f) &= \phi(t(f)) - 2\pi f t(f) + \frac{\pi}{4} \,.
\end{align}
\ees

Applying this to the individual $h_{\ell m}$ modes~\eqref{eq:hlmtd}, treating the $\Delta\phi_{\ell m}$ as constants, defining $\omega_{\ell m} = \dot{\phi}_{\ell m} \simeq m \omega_{\rm orb}$, each mode will have a separate time-to-frequency correspondence
\be
	\omega_{\ell m} \left( t^{\ell m}(f) \right) = 2\pi f \,,
\ee
and the phase
\be\label{eq:PsilmSPA}
	\Psi_{\ell m}^{\rm SPA}(f) = m \phi_{\rm orb}\left(t^{\ell m}(f) \right) - 2\pi f t^{\ell m}(f) + \Delta\phi_{\ell m} + \frac{\pi}{4} \,.
\ee
This is the Fourier-domain equivalent to the time-domain relation~\eqref{eq:philmscaling}. We note useful relations between different mode numbers. The various $t^{\ell m}(f)$ functions are related by
\be
	t^{\ell m}\left(\frac{mf}{2}\right) = t^{22}(f) \,,
\ee
while the phases satisfy
\be\label{eq:PhilmRelationSPA}
	0 = 2 \Psi_{\ell m}\left( \frac{m f}{2} \right) - m \Psi_{22}\left( f\right) - 2 \Delta\phi_{\ell m} + m \Delta\phi_{22} - \left(2 - m\right)\frac{\pi}{4} \,.
\ee
This last relation holds regardless of the time and phase alignment of the waveform: as a phase change $\delta \phi_{\ell m} = m\delta\phi$, or a time change $\delta\Psi_{\ell m} (f) = -2\pi f \delta t$ would both leave $2\Psi_{\ell m}(mf/2) - m \Psi_{22}(f)$ invariant. It is sensitive however to the quantities $\Delta\phi_{\ell m}$ (that we treat here as constants in the early inspiral), that depend on the choice of polarization convention.

Finally, we recall that we can build a time-to-frequency correspondence directly from the Fourier-domain waveform as
\be\label{eq:deftf}
	t(f) \equiv -\frac{1}{2\pi} \frac{d\Psi}{df}.
\ee
Note that this definition of time is strictly speaking only accurate in the inspiral phase, where the SPA applies and the two definitions~\eqref{eq:deftf} and~\eqref{eq:deftfSPA} coincide. However, it can be used as a proxy for time everywhere, as we can evaluate~\eqref{eq:deftf} for any frequency\footnote{The converse is not true: since $t(f)$ is not monotonic at high frequencies, building an unambiguous mapping $f(t)$ is only possible in the inspiral.} $f$.

\subsubsection{Orbital carrier}
\label{sec:carrier}

In order to carry over information about the alignement of the respective mode from the time domain to the Fourier domain, we find it convenient to introduce a fictitious carrier signal $k(t)$, that evolves with the orbital phase instead of twice the orbital phase as the $h_{22}$ mode does.
\be\label{eq:defcarrier}
	k(t) \equiv a_{22}(t) \exp \left[-i \frac{\phi_{22}(t)}{2} \right] \,.
\ee
The choice made here of keeping the same amplitude as the $h_{22}$ mode is quite arbitrary, but will ensure that it decays in the ringdown, giving us a smooth Fourier transform for this carrier. Note that this construction is artificial, as the carrier does not correspond to any physical signal.

As mentioned before, the carrier half-phase $\phi_{22}/2$ comes with a $\pi$-degeneracy. We can alleviate this by forcing the carrier phase to be within $\pi$ of the orbital phase, as read from the SEOB dynamics, at the time of alignment. This is, in fact, our main motivation for building this carrier in the time domain: it allows us to avoid the issues listed above, with all the conditioning being ultimately tied to the orbital phase, a quantity that is smooth across parameter space.

The Fourier transform of the carrier signal introduced in~\eqref{eq:defcarrier} is decomposed as an amplitude and phase as
\be\label{eq:carrierFD}
	\tilde{k}(f) = A_{k}(f) \exp\left[ -i \Psi_{k}(f)\right] \,,
\ee
where $A_{k} = |\tilde{k}|$ will be discarded in the rest of the conditioning. When the SPA applies, we have approximately
\be\label{eq:PsikSPA}
	\Psi_{k}(f) \simeq \Psi_{k}^{\rm SPA}(f) = \phi_{\rm orb}(t^{k}(f)) - 2\pi f t^{k}(f) + \frac{\pi}{4}\,,
\ee
with $t^{k}(f)$ defined like in~\eqref{eq:deftfSPA} as
\be
	\omega_{\rm orb}(t^{k}(f)) = 2\pi f \,.
\ee

Since $\tilde{k}(f)$ is computed via an FFT, nothing forbids arbitrary jumps of $2\pi$ of the phases between different points in parameter space. We use the relation above to remove this $2\pi$-ambiguity in $\Psi_{k}$. At the frequency $f_{\rm align}$ such that $t(f_{\rm align}) = t_{\rm align}$, we impose
\be
	\left| \Psi_{k}(f_{\rm align}) - \left(\phi_{\rm orb}(t_{\rm align}) - 2\pi f_{\rm align} t_{\rm align} + \frac{\pi}{4} \right)\right| < \pi \,.
\ee
In this way, $\Psi_{k}$ is directly tied to $\phi_{\rm orb}$ that is smooth in parameter space in our time-domain conditioning procedure.

We will factor out the Fourier domain phase of the carrier, build a ROM for the carrier separately, and then factor in the modelled carrier phase when reconstructing the waveform.

\subsubsection{Fourier-domain coorbital modes and waveform building blocks}
\label{sec:coorbital_modes_and_waveform_building_blocks}

Next, we build coorbital modes by scaling out the Fourier-domain phase of the carrier following
\be\label{eq:defhcoorb}
	\tilde{h}_{\ell m}^{c}(f) = \tilde{h}_{\ell m}(f) \exp\left[i m \Psi_{k}(f/m)\right] \,.
\ee
These modes are built so as to factor out the main contribution to the phase of the Fourier-domain modes, to leave the coorbital modes $\tilde{h}_{\ell m}^{c}$ with an approximately constant phase in the inspiral regime. Namely, for the inspiral regime, where the SPA is valid, $t^{k}(f) = t^{\ell m}(mf)$ and applying~\eqref{eq:PsilmSPA} and \eqref{eq:PsikSPA} gives
\be\label{eq:PsilmfromPsik}
	\Psi_{\ell m}(f) \simeq m \Psi_{k}\left(\frac{f}{m}\right) + \Delta\phi_{\ell m} + (1 - m) \frac{\pi}{4} \,.
\ee
Note that our Fourier-domain construction is approximate, and these ``coorbital'' quantities $\tilde{h}^{c}_{\ell m}$ do not correspond exactly to a coorbital frame defined in the time domain as in~\cite{Blackman:2017pcm, Varma:2019csw}.

We stress that these modes are not strictly coorbital, in the sense that they are not built from a time-domain coorbital frame built from the orbital phase. Indeed, the definition~\eqref{eq:defhcoorb} is rooted in the Fourier domain, and its physical meaning is unclear in the high-frequency range where the SPA does not apply anymore.

Thus, the basic building blocks for the ROM will be
\begin{itemize}
	\item $\Psi_{k} = - \mathrm{Arg}\left[ \tilde{k}\right]$, the Fourier-domain carrier phase;
	\item $\mathrm{Re}\left( \tilde{h}_{\ell m}^{c} \right)$, the real part of the coorbital modes;
	\item $\mathrm{Im}\left( \tilde{h}_{\ell m}^{c} \right)$, the imaginary part of the coorbital modes.
\end{itemize}
Conversely, to rebuild the modes $\tilde{h}_{\ell m}$ from these waveform pieces, it is enough to factor in the carrier phase as in~\eqref{eq:defhcoorb}.

\subsubsection{Scaling of frequencies using ringdown frequency}
\label{sec:scaling_of_frequencies}

One of the prerequisites of our ROM procedure is to represent the waveform on a common frequency grid. However, the frequency range covered varies with physical parameters, notably spin. In \texttt{SEOBNRv4\_ROM}, this was alleviated by extending waveforms to higher frequencies. Here, we choose to apply a scaling to the frequencies of the waveform building blocks, depending on the ringdown frequency. For every mode $(\ell,m)$ and the carrier we define
\begin{subequations}
\begin{align}
\label{eq:freq_scalings}
	y_{\ell m} &= \frac{2\pi}{\omega^{\rm QNM}_{\ell m}} Mf \,, \\
	y_{k} &= \frac{4\pi}{\omega^{\rm QNM}_{22}} Mf \,,
\end{align}
\end{subequations}
where $\omega^{\rm QNM}_{\ell m}$ is the quasi-normal mode frequency, and varies for different waveforms as it depends on the spin of the remnant black hole. We will then use for all waveforms a common grid of this rescaled parameter $y$. Given this scaling, we have to carefully adjust the starting frequency of the waveforms of our training set so that the frequency range of the carrier phase $\Psi_{k}$ covers all modes after undoing the scaling. The maximal values of $y_{\ell m}$, $y_{k}$ where we cut the data are $(y^{\rm \max}_{22}, y^{\rm \max}_{21}, y^{\rm \max}_{33}, y^{\rm \max}_{44}, y^{\rm \max}_{55}) = (1.7, 1.7, 1.55, 1.35, 1.25)$ and $y^{\rm \max}_{k} = 2.5$. This technique is only used for building the high-frequency ROM (see Sec.~\ref{sec:patching_parameter_space}); for the low-frequency ROM, the ringdown frequency is irrelevant and the scaling would induce an additional cost in generating the waveforms of the training set.

\subsection{SVD decomposition}
\label{sec:SVD}

We decompose all waveform data pieces defined in Sec.~\ref{sec:coorbital_modes_and_waveform_building_blocks} into respective \ac{SVD} bases and subsequently interpolate the projection coefficients in each \ac{SVD} basis over the parameter space, as discussed in Sec.~\ref{sec:TPI}. This method follows earlier work in~\cite{Purrer:2014fza,Purrer:2015tud,Bohe:2016gbl}.

We start with a waveform data piece $X(f_i; \vec\theta)$, given on a discrete grid in frequency $\{f_i\}_{i=1}^m$, and on a regular grid of points $\vec\theta$ in the three-dimensional binary parameter space in mass-ratio 
$q$ 
and aligned \ac{BH} spins 
$\chi_i$, 
$(q, \chi_1, \chi_2)$. We flatten the parameter grid and arrange the data in matrix form $X_{ij} = X(f_i; \theta_j) \in \mathbb{R}^{m \times n}$, where
$n$ is the total number of input waveforms.

We then resample the data in a log-spaced frequency grid of $300$ points. The number of points used for resampling are based on previous studies (see Ref.~\cite{Bohe:2016gbl,Purrer:2015tud}).
We compute the \ac{SVD}~\cite{GolubVanLoan,Demmel} $X = V \Sigma U^T$ and obtain an orthonormal basis for the column space of the matrix $X$ in the first $r = \rank(X)$ columns of $V$. The \ac{SVD} provides us with a decomposition of the range space of $X$, $\range(X) =  \operatorname{span} \{ v_1, \dots, v_r \}$, where the $v_j$ are the \emph{left singular vectors} of $X$.

Given the basis $\mathcal{B}_X = V$, we expand the waveform data pieces $x_j(f_i)$ that make up the columns of $X$ in this basis and can write the expansion $x_j \approx \sum c_X(\theta_j) \cdot \mathcal{B}_X$ with projection coefficient matrix $c_X = \mathcal{B}^T_X \cdot X$.
To construct a waveform model we need to predict the coefficients $c_X$ at a desired parameter space point $\theta^*$. To do that we need to fit or interpolate $c_X$ over the parameter space. This is discussed in Sec.~\ref{sec:TPI}.

\subsection{Tensor-product spline interpolation}
\label{sec:TPI}

In the following we describe how we obtain projection coefficients at arbitrary parameter space points. In low dimensional spaces we can afford to use dense grids built from the Cartesian product of one-dimensional sets of points. We choose cubic splines as the univariate interpolants and obtain a \ac{TPI}~\cite{Purrer:2014fza,githubTPI,Setyawati:2019xzw} for a three-dimensional coefficient tensor $c_{ijk}$ which can be written as
\begin{equation}
  \label{eq:TPI}
  I_\otimes[c](q, \chi_1, \chi_2) = \sum_{ijk} c_{ijk}
    \left( \Psi_i \otimes \Psi_j \otimes \Psi_k \right)(q, \chi_1, \chi_2).
\end{equation}
Here, the $\Psi$ are B-spline basis functions~\cite{deBoor} of order 3 for the chosen one dimensional sets of parameter space points in each dimension. We use ``not-a-knot'' boundary conditions to avoid having to specify derivatives at the domain boundaries.
We built the model using \texttt{TPI}, a \texttt{Cython/C} package~\cite{githubTPI} to provide tensor product spline interpolation in \texttt{Python}, and later implemented the model in  LAL~\cite{LALSuiteGit}.

\subsection{Patching in geometric frequency}
\label{sec:patching-in-frequency}

Here we discuss dividing the waveform domain in geometric frequency into separate sub-domains,  where we build a separate ROMs. In Sec.~\ref{sec:patching_parameter_space} we will instead discuss how to tackle non-uniform resolution requirements over the binary parameter space.

In the early inspiral waveforms modes tend to be well approximated by the PN expansion and an accurate ROM can be built from a relatively small training set.
In contrast, the high geometric frequency part of the waveform modes encapsulates the late inspiral, merger and ringdown part of the signal which is more complex and non-linear, and this is also the regime where EOB waveforms are tuned against NR where they are available. Building an accurate ROM for the high geometric frequency part of the waveform modes consequently requires a higher density of training set waveforms.

Therefore, it is natural to treat the low and high geometric frequency part of the waveform separately, following the construction of previous ROMs~\cite{Purrer:2015tud}. This allows us to make the training set for the low geometric frequency part of the waveforms significantly smaller and reduce the computational cost of the training. Waveforms for low mass binaries are the most costly waveforms to generate. The cost is exacerbated due to the presence of higher modes with $|m|>2$, since they require a lower starting orbital frequency to cover the same gravitational wave frequency range as the dominant mode.

\begin{figure}[H]
        \centering
        \includegraphics[width=0.5\textwidth]{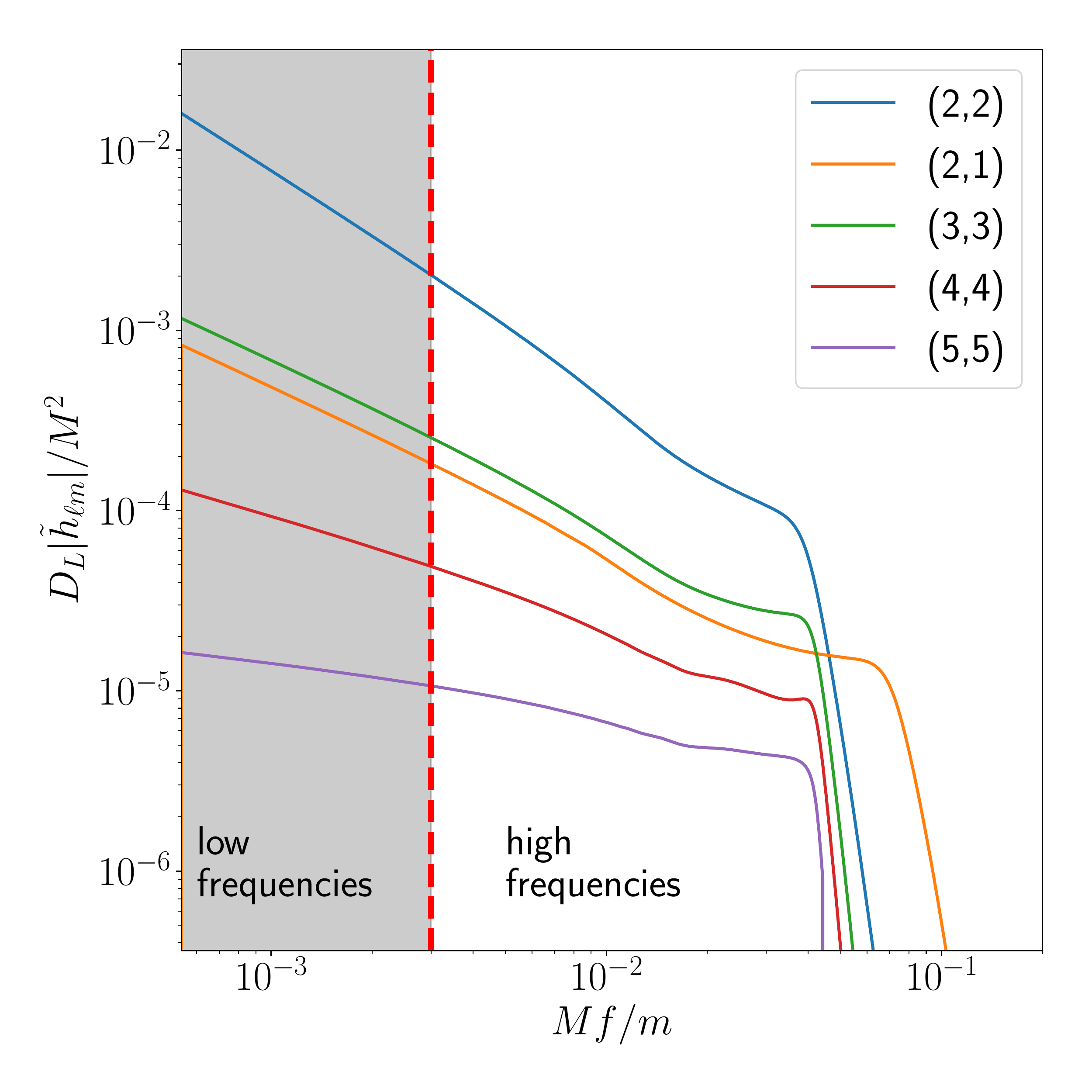}
	\caption{The complete ROM for each waveform mode is build by hybridizing a low and high frequency ROM. The $x$-axis shows the geometric frequency rescaled for each mode $(\ell,m)$ as $M f / m$, following the natural inspiral scaling of the frequency of the waveform modes with $m$.
  The low frequency sub-domain (black shaded region) starts at a geometric frequency of $Mf / m = 0.00025$, and
  transitions to the high frequency sub-domain at $Mf / m = 0.003$.
  }
  \label{fig:frequency-patches}
\end{figure}

In Fig.~\ref{fig:frequency-patches} we show the sub-division into low and high geometric frequency sub-domains. The low frequency sub-domain is connected with PN waveforms modes in the early inspiral,
as discussed in Sec.~\ref{sec:hybridization}. We generated the \texttt{SEOBNRv4HM} waveforms at a sufficiently low frequency (at $15$ Hz and a total mass of $5 M_\odot$ to allow for some tapering) to have the complete set of higher harmonics included in \texttt{SEOBNRv4HM} present at a frequency $Mf = 0.0005 * 5/2 \approx 0.0012$, where the low geometric frequency sub-domain starts. We choose the geometric transition frequency between the low and high frequency sub-domains to be $M f = 0.003*m$, using the natural inspiral scaling of the frequency of the waveform modes with $m$.. 
For the high frequency sub-domain we generated waveforms choosing the starting frequency as described in Sec.~\ref{sec:scaling_of_frequencies}, ensuring the generated waveforms after undoing the scaling~\eqref{eq:freq_scalings} will cover this transition frequency. The complete waveform modes are then generated by blending together the low and high frequency parts at the frequency using a variant of the Planck taper function described in Ref.~\cite{Hinder:2018fsy}.

\subsection{Hybridization with TaylorF2}
\label{sec:hybridization}

Here we describe how we carry out the hybridization of the \ac{ROM} waveform with the \texttt{TaylorF2} inspiral waveform.

After evaluating the \ac{ROM} waveform for all modes, we generate the \texttt{TaylorF2} amplitude and phase for the $(2,2)$ mode from the lowest frequency necessary to be able to start all inspiral modes at a user specified frequency. We blend the \texttt{TaylorF2} and \ac{ROM} amplitude and phase for the $(2,2)$ mode using the same Planck taper function used to connect high and low frequency ROM. We can obtain the higher mode \ac{PN} inspiral waveforms by rescaling the $(2,2)$ amplitude and phase.
For the phase we follow Eq.~\eqref{eq:PsilmfromPsik} 
and Eq.~\eqref{eq:delta_phi_constants} to compute the carrier phase from the \texttt{TaylorF2} (2,2) phase and rescale it to obtain the phase for each mode. We then align the inspiral phase with the \ac{ROM} phase for each mode and blend them together on a common frequency grid. For the amplitude we rescale the \texttt{TaylorF2} (2,2) amplitude according to the PN amplitudes given in Ref.~\cite{Mishra:2016whh} Eqs.(12a-12t).

\subsection{Patching in parameter space}
\label{sec:patching_parameter_space}

As already noted for the previous ROMs of EOBNR waveforms (see Refs.~\cite{Purrer:2014fza,Purrer:2015tud,Bohe:2016gbl}) model features often require more resolution in particular parts of the parameter space. However, regular grids do not allow for local refinement in $(q, \chi_1, \chi_2)$. Therefore, we partition the binary parameter space into subdomains on which resolution requirements can be satisfied with a particular regular grid choice.

\begin{figure}[H]
        \centering
        \includegraphics[width=0.5\textwidth]{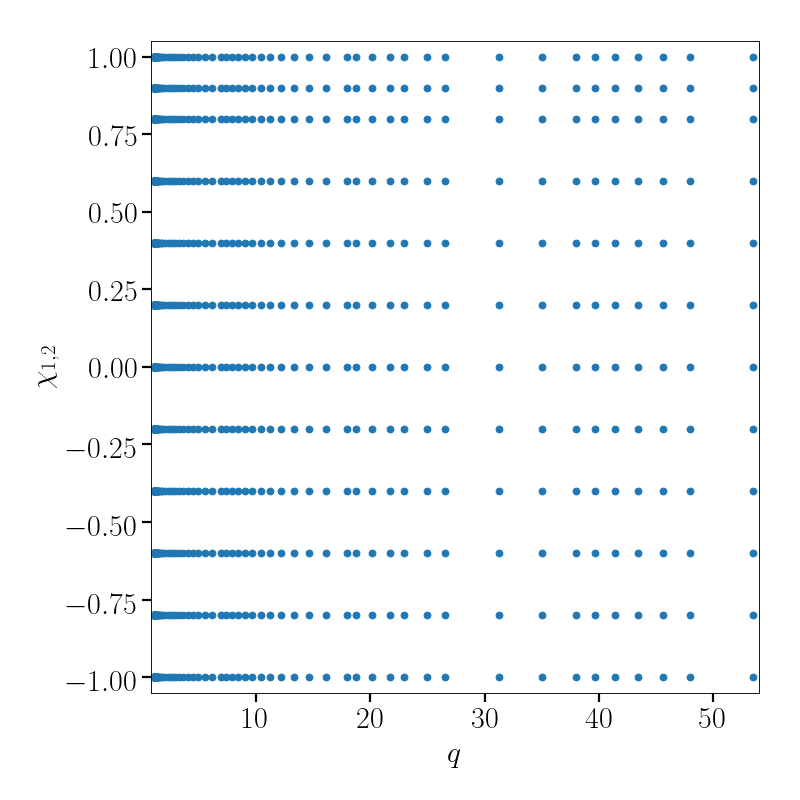}
	\caption{Location in parameter space $(q, \chi_1, \chi_2)$ of the waveforms used to build the inspiral ROM. For this ROM both spin components use the same grid.}
	\label{fig:LFGrid}
\end{figure}

The low frequency ROM does not need any special treatment and was built using waveforms placed on a Cartesian grid with 64 points in $q$ and 12 points in $\chi_{1,2}$ as shown in Fig.~\ref{fig:LFGrid}. Here, the 1D grids in $\chi_1$ and $\chi_2$ were chosen to be identical. The grids for $q$ and $\chi_{1,2}$ are the same as the ones used for \texttt{SEOBNRv4\_ROM} (see Sec.VII in Ref.~\cite{Bohe:2016gbl}), except that we limit the grid to $q = 50$.

On the other hand, as already noted in Refs.~\cite{Purrer:2015tud,Bohe:2016gbl}, modeling the non-linear merger and ringdown part of the waveform in the high geometric frequency ROM requires a higher resolution when approaching large positive values of the primary spin. Therefore, we build two different high frequency ROMs based on the value of the primary's spin, with one ROM having a finer grid in the region of high $\chi_1$.
The inclusion of higher modes in the \texttt{SEOBNRv4HM\_ROM} model also requires additional resolution near equal mass. The modes with odd $m$ vanish by symmetry on the line $q = 1$ and $\chi_1 = \chi_2$ and their behavior in the vicinity is non-trivial to model (see Refs.~\cite{Cotesta:2018fcv,Nagar:2020pcj}). Therefore, we build two different high geometric frequency ROMs in mass-ratio, one of which is covering the region $q \to 1$ with a finer grid. In total we then have four high frequency ROMs to correctly model the merger and ringdown part of the signal.

\begin{figure}[H]
        \centering
        \includegraphics[width=0.5\textwidth]{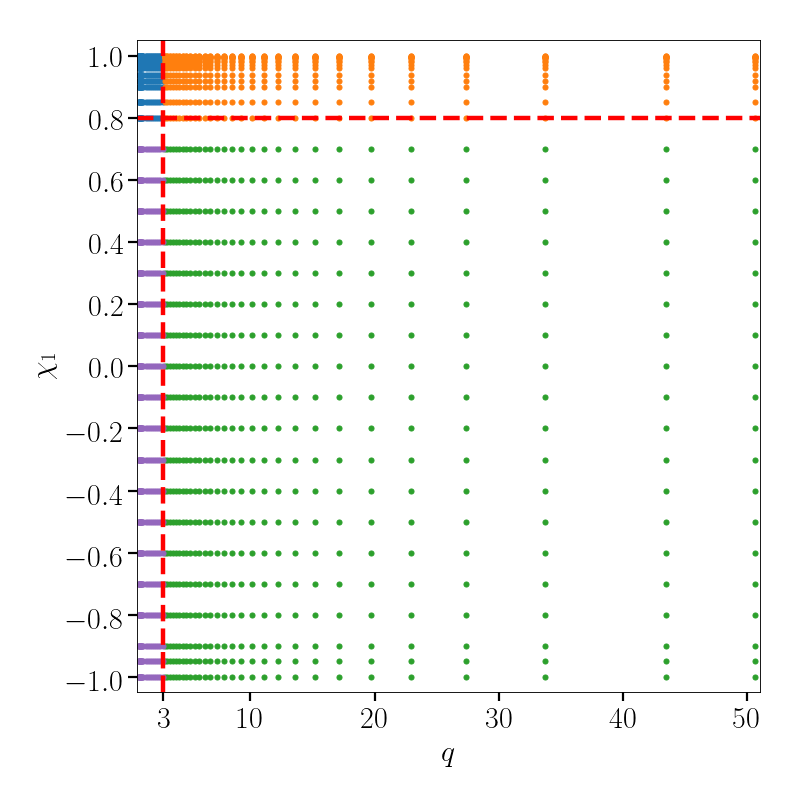}
	\caption{Location in the parameter space $(q, \chi_1)$ of the waveforms used to build the high-frequency ROM. The dashed red lines in the plot are the boundaries of the different patches, at $\chi_1 = 0.8$ and $q = 3$.
  }
\label{fig:HFGrid}

\end{figure}

The 2D projection of the grid in $(q, \chi_1)$ for these four ROMs is shown in Fig.~\ref{fig:HFGrid}. Since no special choice is made for the grid in $\chi_2$ we have omitted plotting the grid in this dimension. We set domain boundaries at $q = 3$ and $\chi_1 = 0.8$ for these four ROMs. In Table~\ref{tab:subdomains} we collect information on how the four patches are placed in parameter space and the number of gridpoints in each dimension.

\begin{table}
 \begin{tabular}{c|lclcl|c}
\hline
\hline
Patch    & \multicolumn{5}{c}{Intervals in $(q, \chi_1, \chi_2)$} & Points per interval\\
\hline
Patch 1  &  $[1,3]$  & $\cup$ & $[-1, 0.8]$ & $\cup$ & $[-1,1]$ & $24 \times 24 \times 24$\\
Patch 2  &  $[1,3]$  & $\cup$ & $(0.8, 1] $ & $\cup$ & $[-1,1]$ & $24 \times 17 \times 24$\\
Patch 3  &  $[3,50]$ & $\cup$ & $[-1, 0.8]$ & $\cup$ & $[-1,1]$ & $31 \times 24 \times 24$\\
Patch 4  &  $[3,50]$ & $\cup$ & $[0.8,1]  $ & $\cup$ & $[-1,1]$ & $30 \times 19 \times 24$\\
\hline
\hline
\end{tabular}
\caption{
  The grids for the high frequency \acp{ROM} on the four patches in parameter space shown in Fig.~\ref{fig:HFGrid}. The physical domain covered by each patch is defined by a Cartesian product of intervals in binary parameters 
  $(q, \chi_1, \chi_2)$. We also indicate the number of grid points in each parameter per patch.
}
\label{tab:subdomains}
\end{table}

\section{Results}
\label{sec:results}

In this section we discuss the accuracy and the increase in efficiency of \texttt{SEOBNRv4HM\_ROM} compared to \texttt{SEOBNRv4HM}.
Finally we also perform a parameter estimation study to demonstrate the potential of this model in data analysis applications.

\subsection{Accuracy of the model}
\label{sec:accuracy}

We start by defining the faithfulness function, used to assess the closeness between two waveforms when higher-order modes are included. We then use this faithfulness measure to determine how accurately the ROM reproduces \texttt{SEOBNRv4HM} waveforms.
\subsubsection{Definition of faithfulness}

A GW signal emitted by a spinning, non-precessing and non-eccentric
BBH is characterized by 11 parameters in the detector frame. These parameters are the BH masses $m_{1}$ and $m_{2}$, the (constant) projection
of the spins in the
direction perpendicular to the orbital plane $\chi_{1}$ and $\chi_{2}$, the angular position of the line of sight measured in the source's
frame $(\iota, \varphi_0)$,
the sky location of the source in the detector frame $(\theta, \phi)$, the luminosity distance $D_{\mathrm{L}}$,
the time of arrival $t_{\mathrm{c}}$ of the signal and finally the
polarization angle $\psi$.
The detector response can be written as:
\begin{align}
\label{eq:det_strain}
h \equiv & F_+(\theta,\phi,\psi) \ h_+(\iota,\varphi_0, D_{\mathrm{L}}, \boldsymbol{\xi},t_{\mathrm{c}};t) \nonumber \\
&+ F_\times(\theta,\phi,\psi)\ h_\times(\iota,\varphi_0, D_{\mathrm{L}}, \boldsymbol{\xi},t_\mathrm{c};t)\,,
\end{align}
where masses and spins are combined in the vector  $\boldsymbol{\xi} \equiv (m_{1}, m_{2}, \chi_{1}, \chi_{2})$, and the functions $F_+(\theta,\phi,\psi)$ and $F_\times(\theta,\phi,\psi)$ are the antenna patterns~\cite{Sathyaprakash:1991mt,Finn:1992xs}.
This equation can be rewritten as:
\begin{align}
h = & \mathcal{A}(\theta,\phi)\big[\cos\kappa(\theta,\phi,\psi) \ h_+(\iota, \varphi_0, D_{\mathrm{L}}, \boldsymbol{\xi}, t_{\mathrm{c}};t) \nonumber \\
&+ \sin\kappa(\theta,\phi,\psi) \ h_\times (\iota, \phi, D_{\mathrm{L}}, \boldsymbol{\xi},t_{\mathrm{c}};t) \big],
\end{align}
with $\kappa(\theta,\phi,\psi)$ being the \textit{effective polarization}~\cite{Capano:2013raa} defined in the range $[0, 2\pi)$ as:
\begin{equation}
e^{i \kappa(\theta,\phi,\psi)} = \frac{F_+(\theta,\phi,\psi) + i F_\times(\theta,\phi,\psi)}{\sqrt{F_+^2(\theta,\phi,\psi) + F_\times^2(\theta,\phi,\psi)}},
\end{equation}
where the function $\mathcal{A}(\theta,\phi)$ is an overall amplitude and is defined as:
\begin{equation}
\mathcal{A}(\theta,\phi) = \sqrt{F_+^2(\theta,\phi,\psi) + F_\times^2(\theta,\phi,\psi)}\,.
\end{equation}
 Given a GW signal $h_{\mathrm{s}}$ (\texttt{SEOBNRv4HM} in our case) and a template waveform $h_{\mathrm{t}}$ (\texttt{SEOBNRv4HM\textunderscore ROM} in this context), we define the faithfulness (or match) as~\cite{Capano:2013raa,Harry:2016ijz,Cotesta:2018fcv}:
\begin{equation}
\label{eq:faith}
\mathcal{F}(\iota_{\textrm{s}},{\varphi_0}_{\textrm{s}},\kappa_{\textrm{s}}) \equiv  \max_{t_c, {\varphi_0}_{\mathrm{t}}, \kappa_{\textrm{t}}} \left[\left . \frac{ \left( h_{\mathrm{s}},\,h_{\mathrm{t}} \right)}{\sqrt{ \left( h_{\mathrm{s}},\,h_{\mathrm{s}} \right) \left( h_{\mathrm{t}},\,h_{\mathrm{t}} \right)}}\right \vert_{\substack{\iota_{\mathrm{s}} = \iota_{\mathrm{t}} \\\boldsymbol{\xi}_{\mathrm{s}} = \boldsymbol{\xi}_{\mathrm{t}}}} \right ],
\end{equation}
where parameters with the subscript ``s'' (``t'') refer to the signal (template) waveform. The expression above does not depend on $\mathcal{A}(\theta,\phi)$, therefore the only dependance on $(\theta,\phi,\psi)$ is encoded in $\kappa(\theta,\phi,\psi)$.
For the faithfulness calculation we optimize over the phases ${\varphi_0}_{\mathrm{t}}$ and $\kappa_{\textrm{t}}$ and the time of arrival $t_c$ because they are not interesting from an astrophysical perspective.
We use the standard definition of the inner product between two waveforms (see~\cite{Sathyaprakash:1991mt,Finn:1992xs}):
\begin{equation}
\left( a, b\right) \equiv 4\ \textrm{Re}\int_{f_\textrm{low}}^{f_\textrm{high}} df\,\frac{\tilde{a}(f) \ \tilde{b}^*(f)}{S_n(f)},
\end{equation}
where \char`\~ \, denotes the Fourier transform, * indicates the complex
conjugate and $S_n(f)$ is the one-sided power spectral density (PSD)
of the detector noise. For this computation we use the Advanced LIGO ``zero-detuned
high-power'' design sensitivity curve~\cite{Barsotti:2018}. The integral is computed between the
frequencies $f_{\textrm{low}} = 20 \mathrm{Hz}$ and $f_{\textrm{high}} = 3
\mathrm{kHz}$.
The same definition of faithfulness has been used to determine the agreement between \texttt{SEOBNRv4HM} and numerical relativity waveforms (see ~\cite{Cotesta:2018fcv})\footnote{In Ref.~\cite{Cotesta:2018fcv} we used an old version of the Advanced LIGO ``zero-detuned
high-power'' design sensitivity curve (in Ref.~\cite{Shoemaker:2010}). We have checked that the difference in the faithfulness calculation when using the new version of the sensitivity curve was negligible. For this reason here we report the calculations performed with the new curve described in Ref.~\cite{Barsotti:2018}.}.
Since the faithfulness given in Eq.~\eqref{eq:faith} depends on the signal parameters
$(\iota_{\textrm{s}},{\varphi_0}_{\textrm{s}},\kappa_{\textrm{s}})$, we will summarize the results using
the maximum and the average \textit{unfaithfulness} (or mismatch)
$[1-\mathcal{F}(\iota_{\textrm{s}},{\varphi_0}_{\textrm{s}},\kappa_{\textrm{s}})]$
over these parameters, namely~\cite{Buonanno:2002fy,Capano:2013raa,Harry:2016ijz}:

\begin{widetext}
\begin{equation}
\mathcal{U}_{\mathrm{max}} \equiv \max_{\iota_{\mathrm{s}},{\varphi_0}_{\mathrm{s}},\kappa_{\mathrm{s}}}(1 -\mathcal{F}) \equiv  1 - \min_{\iota_{\mathrm{s}},{\varphi_0}_{\mathrm{s}},\kappa_{\mathrm{s}}}\mathcal{F}(\iota_{\textrm{s}},{\varphi_0}_{\textrm{s}},\kappa_{\textrm{s}}) \label{eq:max_unfaith}\,,
\end{equation}
\begin{equation}
\bar{\mathcal{U}} \equiv \langle
1-\mathcal{F}\rangle_{\iota_{\mathrm{s}},{\varphi_0}_{\mathrm{s}},\kappa_{\mathrm{s}}} \equiv  1 - \frac{1}{8\pi^2}\int_{0}^{2\pi} d\kappa_{\mathrm{s}} \int_{-1}^{1} d(\cos\iota_s) \int_{0}^{2\pi} d{\varphi_0}_{\mathrm{s}} \ \mathcal{F}(\iota_{\textrm{s}},{\varphi_0}_{\textrm{s}},\kappa_{\textrm{s}})\,. \label{eq:avg_unfaith}
\end{equation}
\end{widetext}

\subsubsection{Faithfulness against \texttt{SEOBNRv4HM}}

In order to avoid biases in data analysis applications when using the ROM instead of \texttt{SEOBNRv4HM},
it is important to verify that the additional modeling error introduced in the construction of the ROM is negligible compared to the inaccuracy of the \texttt{SEOBNRv4HM} waveforms with respect to the NR simulations.
Since the typical unfaithfulness between \texttt{SEOBNRv4HM} and NR waveforms is $\mathcal{O}(1\%)$ (see Figs. (11) and (12) in Ref.\cite{Cotesta:2018fcv}), it is therefore natural to require
the unfaithfulness between \texttt{SEOBNRv4HM} and \texttt{SEOBNRv4HM\textunderscore ROM} to be $\mathcal{O}(0.1\%)$ or less.
To that end we have generated 10000 \texttt{SEOBNRv4HM} waveforms with random (uniformly distributed) values of $(q, \chi_1, \chi_2)$ and computed their match against the same waveforms produced with \texttt{SEOBNRv4HM\_ROM}.

\begin{figure}[H]
        \centering
        \includegraphics[width=0.5\textwidth]{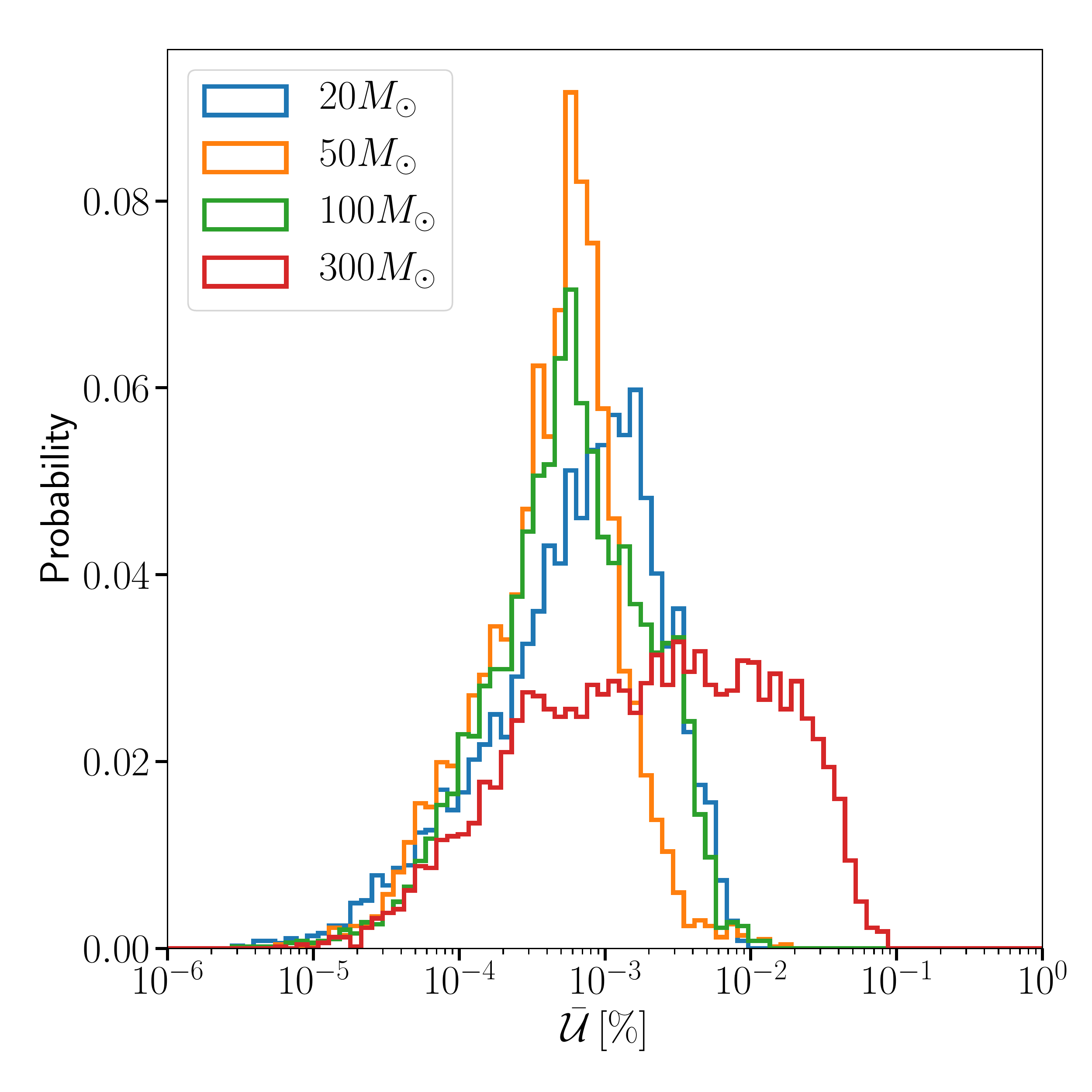}
	\caption{Histograms of the unfaithfulness $\bar{\mathcal{U}}$ (in percent) between \texttt{SEOBNRv4HM} and ROM waveforms for different values of the total mass.
  }
\label{fig:average_unfaith_M}
\end{figure}

We summarize these results in Fig.~\ref{fig:average_unfaith_M} where we show a histogram with the unfaithfulness $\bar{\mathcal{U}}$ computed between the ROM and \texttt{SEOBNRv4HM} waveforms for different values of the total mass. For each total mass we report in Tab.~\ref{tab:average_unfaith_M_summary_table} the median and maximum values of these unfaithfulness distributions.

\begin{table}
 \begin{tabular}{c|c|c|c|c}
\hline
\hline
Total mass $[M_\odot]$             & 20 & 50 & 100 & 300\\
\hline
$\underset{(q, \chi_1,\chi_2)}{\textrm{med}} \bar{\mathcal{U}} [\%]$  & 0.001 & 0.001 & 0.001 & 0.002   \\
\hline
$\underset{(q, \chi_1,\chi_2)}{\max} \bar{\mathcal{U}} [\%]$  & 0.01 & 0.02 & 0.01 & 0.08   \\
\hline
\hline
\end{tabular}
\caption{
  Median and maximum values of the $\bar{\mathcal{U}}$ distributions in
  Fig.~\ref{fig:average_unfaith_M} for different values of the total mass.
}
\label{tab:average_unfaith_M_summary_table}
\end{table}

\begin{figure*}
    \centering
    \begin{minipage}{0.5\textwidth}
        \centering
        \includegraphics[width=1.0\textwidth]{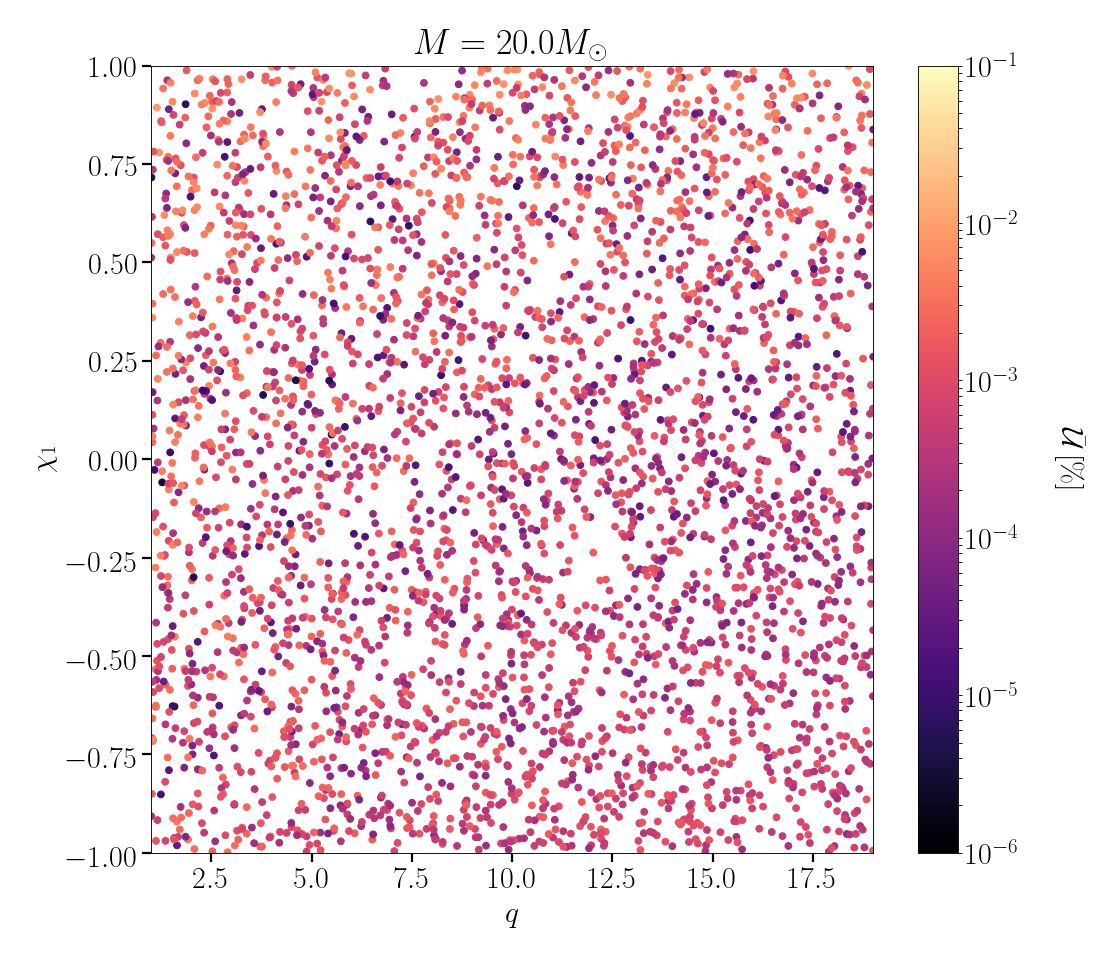}
    \end{minipage}\hfill
    \begin{minipage}{0.5\textwidth}
        \centering
        \includegraphics[width=1.0\textwidth]{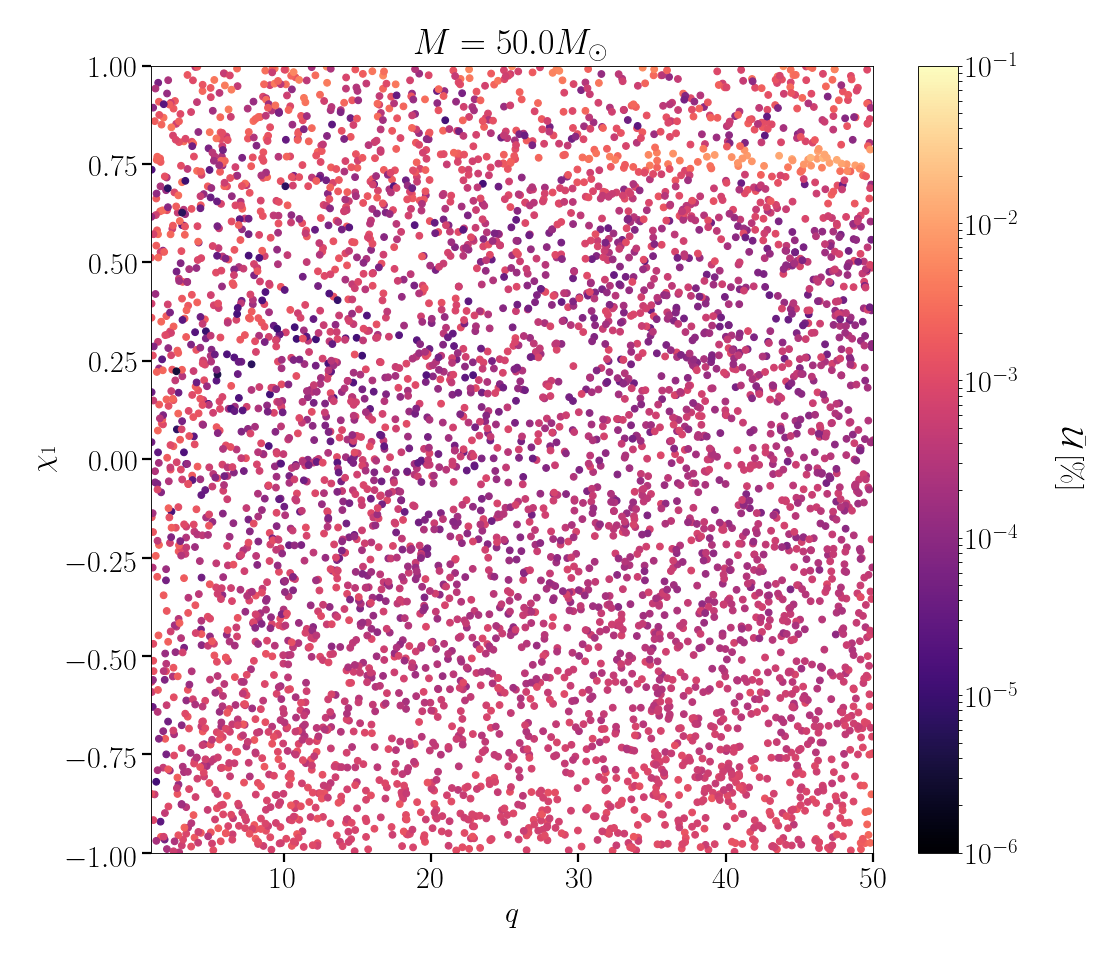}
    \end{minipage}
    \begin{minipage}{0.5\textwidth}
        \centering
        \includegraphics[width=1.0\textwidth]{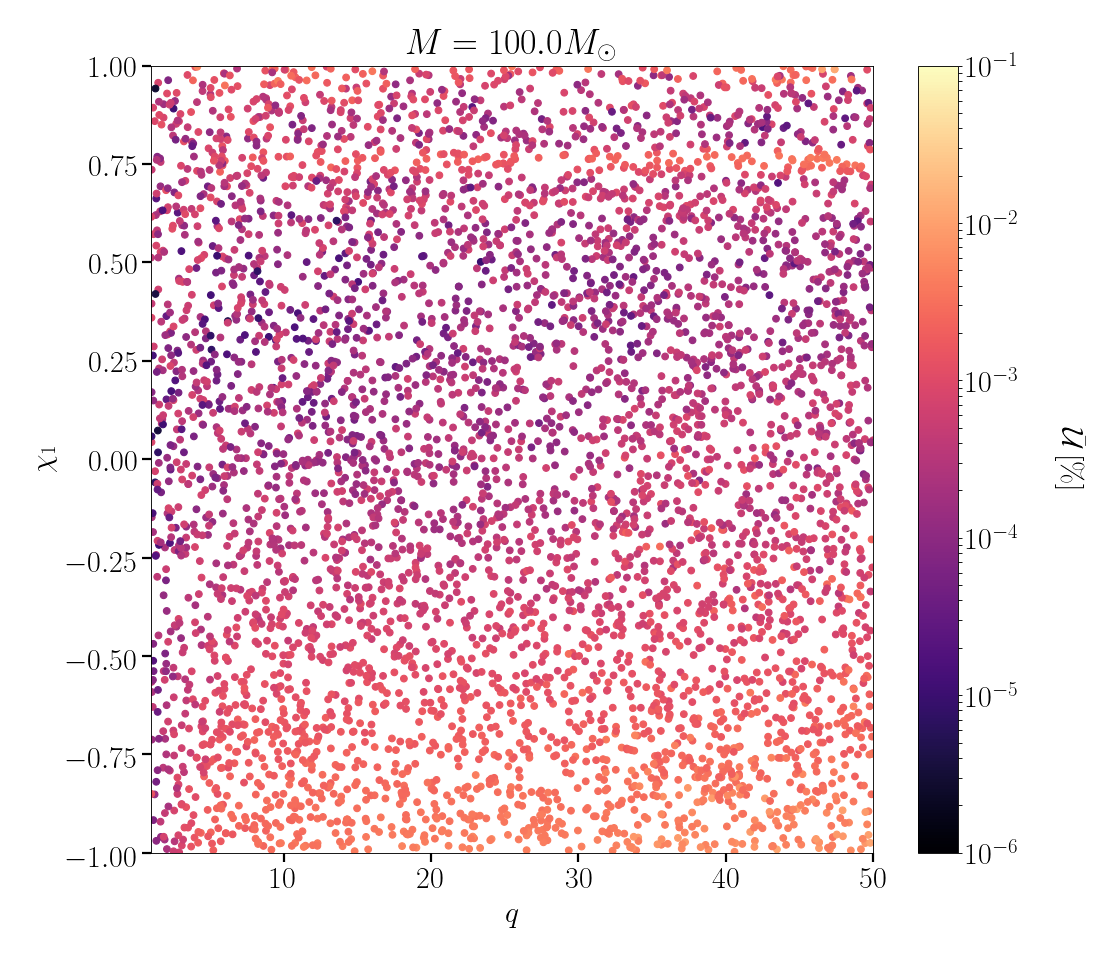}
    \end{minipage}\hfill
    \begin{minipage}{0.5\textwidth}
        \centering
        \includegraphics[width=1.0\textwidth]{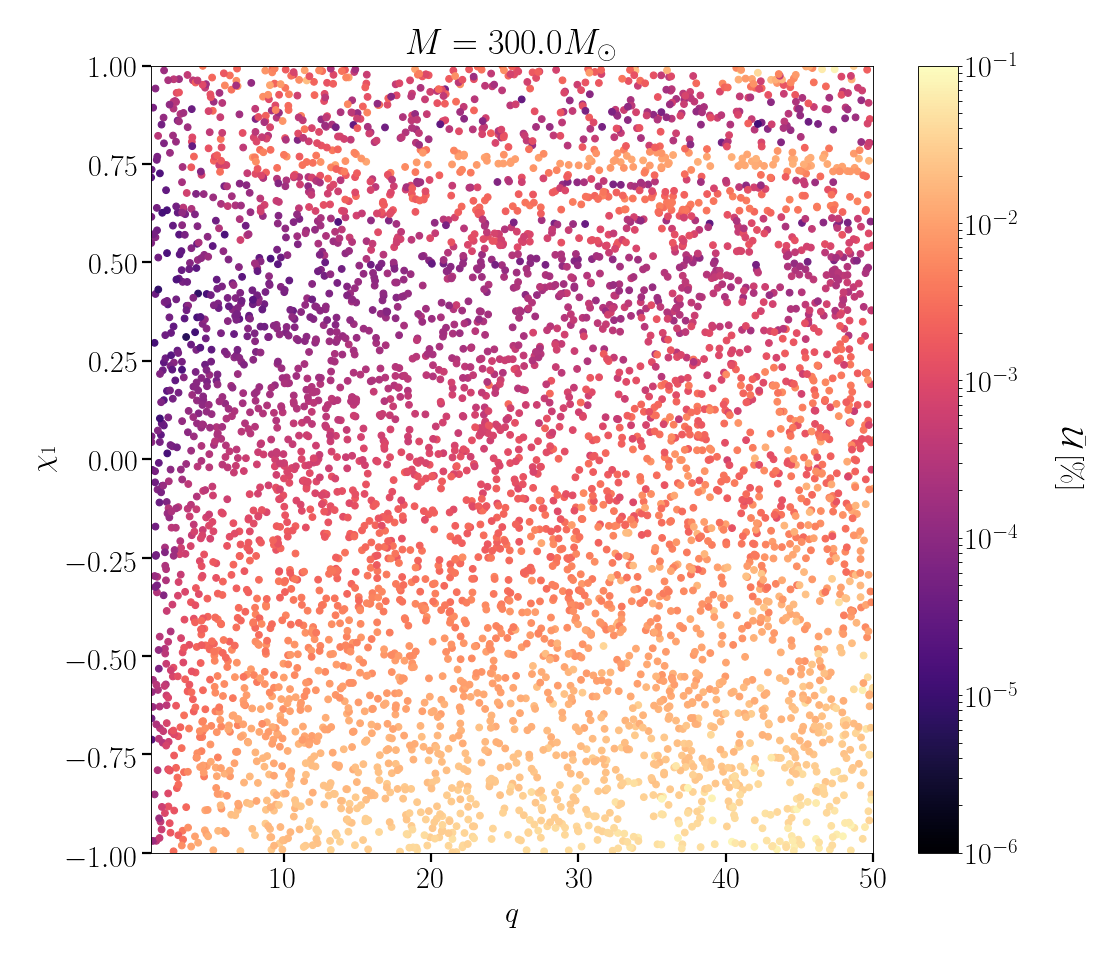}
    \end{minipage}
	\caption{Unfaithfulness $\bar{\mathcal{U}}$ between the ROM and \texttt{SEOBNRv4HM} as a function of $(q,\chi_1)$ and for different values of the total mass.  For $M = 20 M_\odot$ there are no data in the region $q > 19$ because for these system $m_2$ would have an unphysical subsolar mass.}
\label{fig:skyaverageall}
\end{figure*}

The median of these mismatch distributions is weakly dependent on the total mass and it is always $\leq 0.002\%$ while their maximum value is always $\leq 0.08 \%$.
In Fig.~\ref{fig:skyaverageall} we display the distribution of mismatches shown in Fig.~\ref{fig:average_unfaith_M} as a function of $(q, \chi_1)$ and for different values of the total mass. The largest mismatches between the ROM and \texttt{SEOBNRv4HM} are obtained for $M = 300.0 M_\odot$ and large negative $\chi_1$, as it is clear from Fig.~\ref{fig:average_unfaith_M} and Fig.~\ref{fig:skyaverageall} (bottom right panel). ROM GW modes are generated up to a maximum frequency (in geometric units) that scales with the inverse of the total mass of the system. For large total masses the lack of signal above this maximum frequency is the main source of inaccuracy of the ROM. This maximum frequency for each mode is proportional to its least damped quasi-normal mode frequency as described in Eq.~\ref{eq:freq_scalings}. The mismatch is larger for large negative spins because the least damped quasi-normal mode frequency decreases in this region of the parameter space. We highlight that in this region the ROM waveforms still have mismatches $\lesssim 0.1\%$ against \texttt{SEOBNRv4HM} waveforms as demanded at the beginning of this section.
The results described above do not change substantially when considering the distribution of $\mathcal{U}_{\mathrm{max}}$ instead of $\bar{\mathcal{U}}$. In Tab.~\ref{tab:maximum_unfaith_M_summary_table} we report the median and maximum values of these distributions.

\begin{table}
 \begin{tabular}{c|c|c|c|c}
\hline
\hline
Total mass $[M_\odot]$             & 20 & 50 & 100 & 300\\
\hline
$\underset{(q, \chi_1,\chi_2)}{\textrm{med}}\mathcal{U}_{\mathrm{max}} [\%]$  & 0.001 & 0.001 & 0.001 & 0.004   \\
\hline
$\underset{(q, \chi_1,\chi_2)}{\max} \mathcal{U}_{\mathrm{max}}[\%]$  & 0.01 & 0.02 & 0.03 & 0.17   \\
\hline
\hline
\end{tabular}
\caption{
  Median and maximum values of the $\mathcal{U}_{\mathrm{max}}$ distributions for different values of the total mass.
}
\label{tab:maximum_unfaith_M_summary_table}
\end{table}

\begin{figure}[htp]
        \centering
        \includegraphics[width=0.5\textwidth]{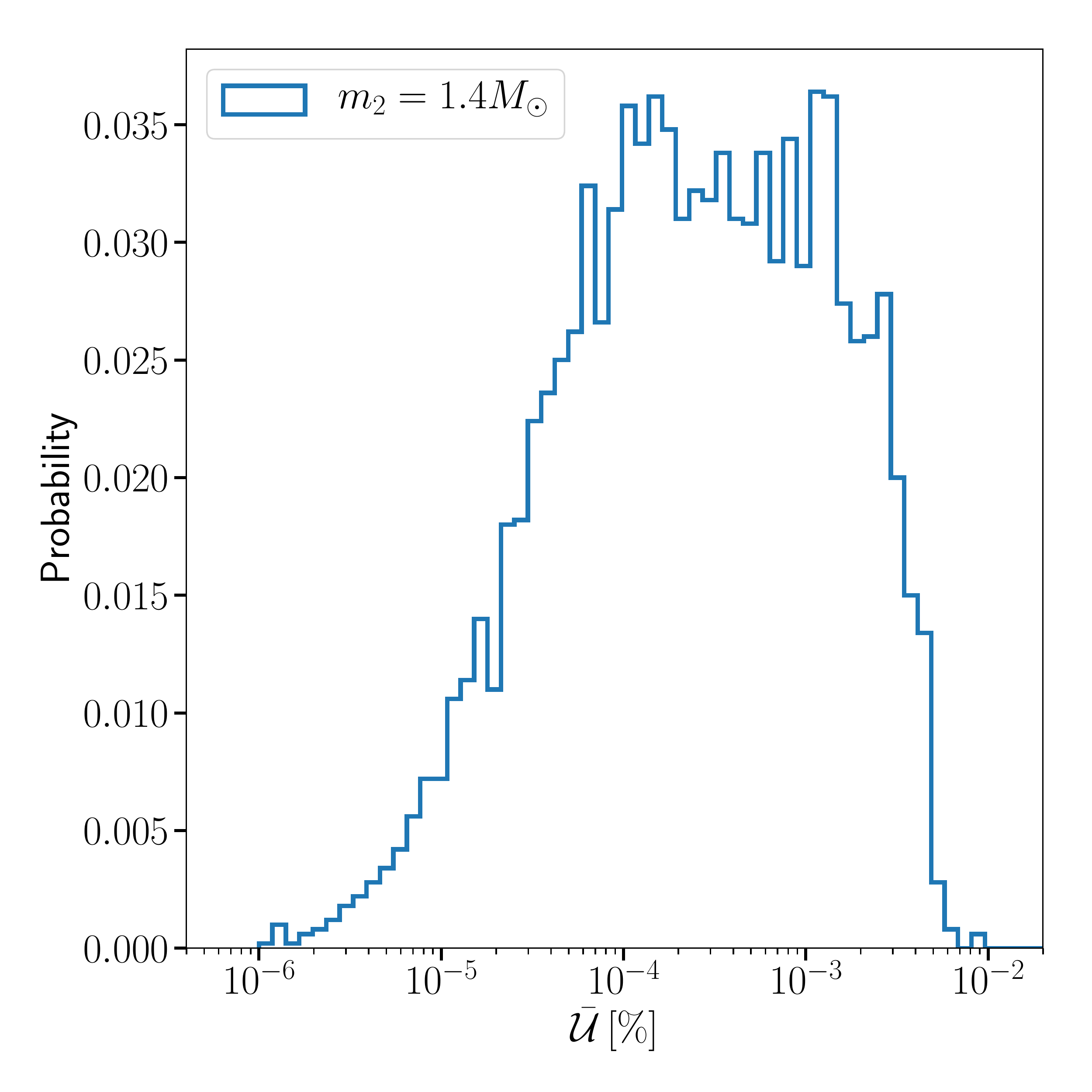}
	\caption{Histogram of the unfaithfulness $\bar{\mathcal{U}}$ between the ROM and \texttt{SEOBNRv4HM}. The \texttt{SEOBNRv4HM} waveforms used in the match calculations have $m_2$ fixed to $1.4 M_\odot$ and $m_1$ uniformly distributed in the range $1.4 M_\odot \leq m_1 \leq 18.6 M_\odot$.}
\label{fig:average_unfaith_m2}
\end{figure}

\begin{figure}[htp]
        \centering
        \includegraphics[width=0.5\textwidth]{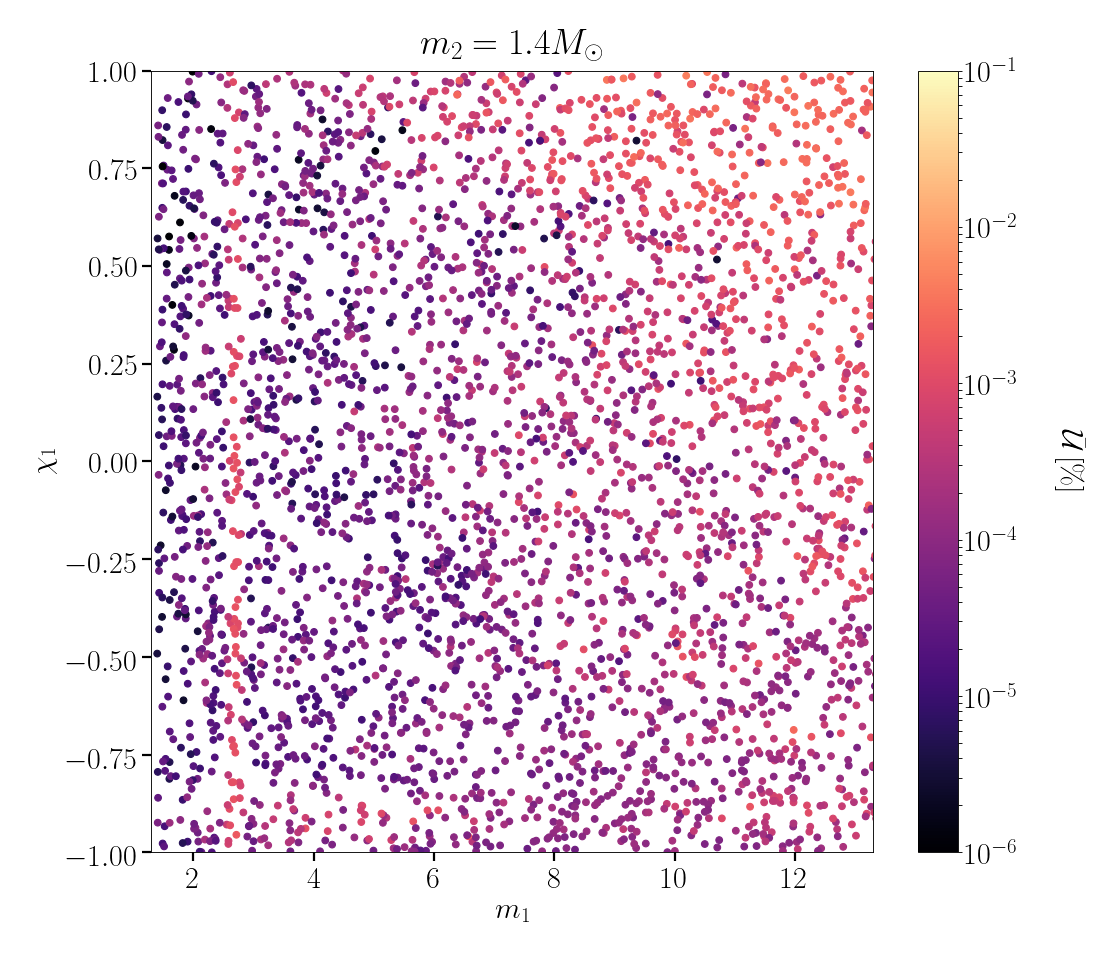}
	\caption{Unfaithfulness $\bar{\mathcal{U}}$ between the ROM and SEOBNRv4HM as a function of $(m_1,\chi_1)$ and with $m_2 = 1.4 M_\odot$.}
\label{fig:skyaverageallm2}
\end{figure}

For total masses $M \leq 20 M_\odot$ it is more convenient to summarize the results of the faithfulness calculations as an histogram with a fixed $m_2$ instead of the total mass.
In Fig.~\ref{fig:average_unfaith_m2} we show the $\bar{\mathcal{U}}$ distribution when fixing  $m_2 = 1.4 M_\odot$ and varying $m_1$ in the interval $1.4 M_\odot \leq m_1 \leq 18.6 M_\odot$ such that the total mass of the system is always $M \leq 20 M_\odot$.
The median of this distribution is $0.0003\%$ while its maximum is $0.01 \%$. In Fig.~\ref{fig:skyaverageallm2} we report the $\bar{\mathcal{U}}$ distribution in Fig.~\ref{fig:average_unfaith_m2} as a function of $(m_1,\chi_1)$. The accuracy of the ROM in this case degrades for large values of $m_1$ and large positive spins but it is still well within the requirements.  Also in this case the results are not very different when looking at the $\mathcal{U}_{\mathrm{max}}$ distribution for which the median is still $0.0003\%$ while the maximum increases to $0.02 \%$.

These analyses demonstrate that the modeling error introduced by the ROM is negligible with respect to the difference between \texttt{SEOBNRv4HM} and NR waveforms. For this reason the mismatch of the ROM against the NR waveforms is essentially the same as \texttt{SEOBNRv4HM} (see Figs.11-12 in Ref~\cite{Cotesta:2018fcv} and Fig.6 in Ref.~\cite{Varma:2018mmi}\footnote{The NR surrogate \texttt{NRHybSur3dq8} has a typical unfaithfulness against the NR simulations of $\mathcal{O}(10^{-3}\%)$, that is neglibigle with respect to the unfaithfulness between the NR simulations and the model \texttt{SEOBNRv4HM} (that is of $\mathcal{O}(1\%)$). Therefore in this case we can consider the \texttt{NRHybSur3dq8} waveform equivalent to an NR waveform. We make the same assumption in the parameter estimation study in Sec.~\ref{sec:PEsec}.}).

\subsection{Computational performance}
\label{sec:computational_performance}

In this section we discuss the computational performance of the ROM in terms of walltime for generating a waveform. We first compare the ROM to \texttt{SEOBNRv4HM} and then to other waveform models that include higher-order modes.

\subsubsection{Speedup with respect to \texttt{SEOBNRv4HM}}
\label{sec:speedup}

The speedup of the ROM with respect to \texttt{SEOBNRv4HM} is computed by the ratio of the walltimes of the two models for generating a frequency domain waveform at the same parameters. Since \texttt{SEOBNRv4HM} is a time domain model, we first generate the waveform in the time domain at a sample rate of 16384 Hz
and then compute its Fourier transform. The ROM waveform is already in the frequency domain and it is generated using the sampling interval set to $1/T$ where $T$ is the duration in seconds of the associated time domain waveform. The maximum frequency of the \texttt{SEOBNRv4HM\_ROM} waveform is set to 8192 Hz.

\begin{figure}[htp]
        \centering
        \includegraphics[width=0.5\textwidth]{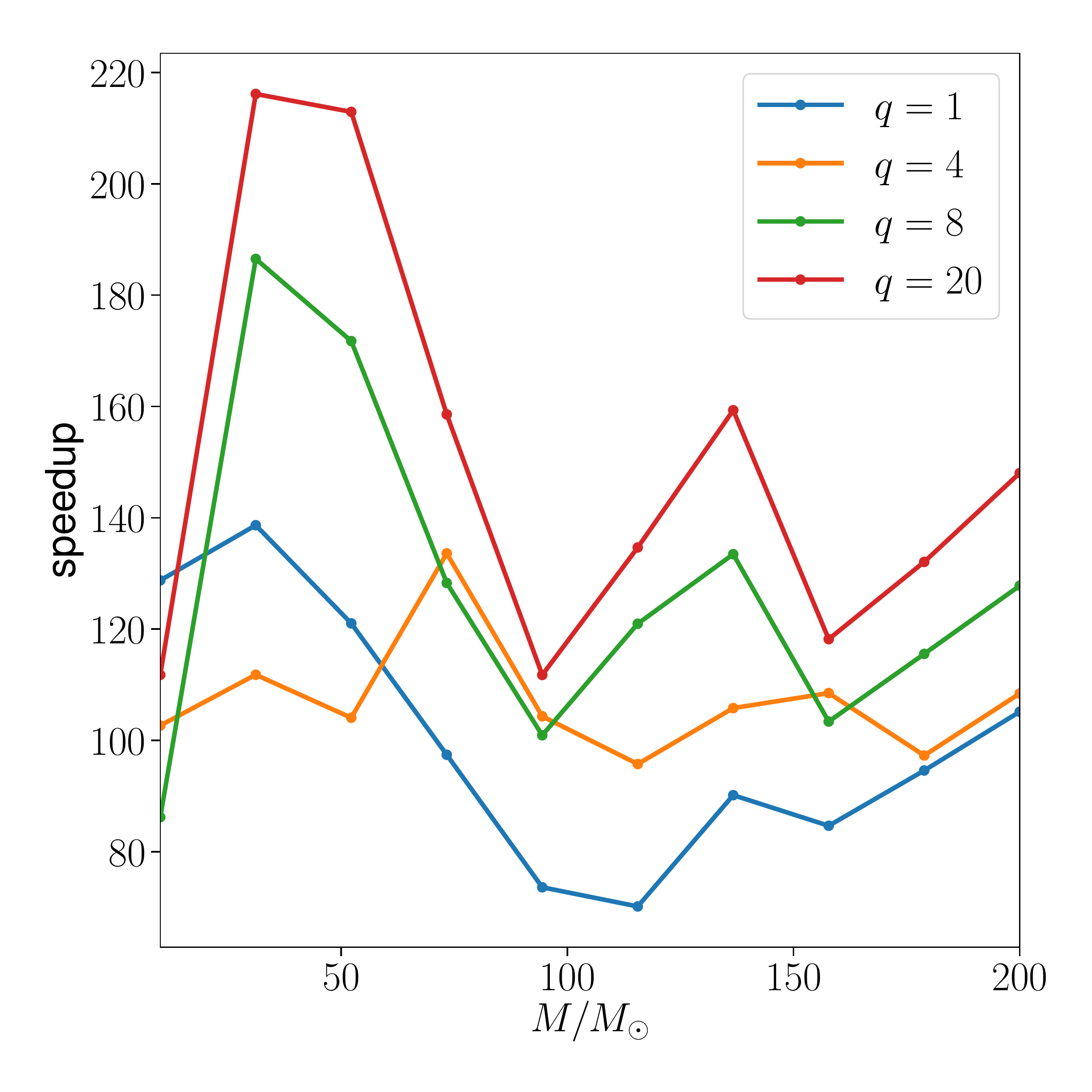}
	\caption{Speedup of waveform generation of the ROM with respect to \texttt{SEOBNRv4HM} as a function of the total mass and for different values of mass ratio.}
	\label{fig:speedup}
\end{figure}

In Fig.~\ref{fig:speedup} we show this speedup as a function of the total mass and for different values of the mass ratio. The speedup is of order $100$. It increases with mass ratio and decreases with total mass. The maximum speedup is found around a total mass of $50 M_\odot$.
Since the spins have only a limited effect on the waveform duration, the speedup depends only weakly on them.

\begin{figure}[htp]
         \centering
         \includegraphics[width=0.5\textwidth]{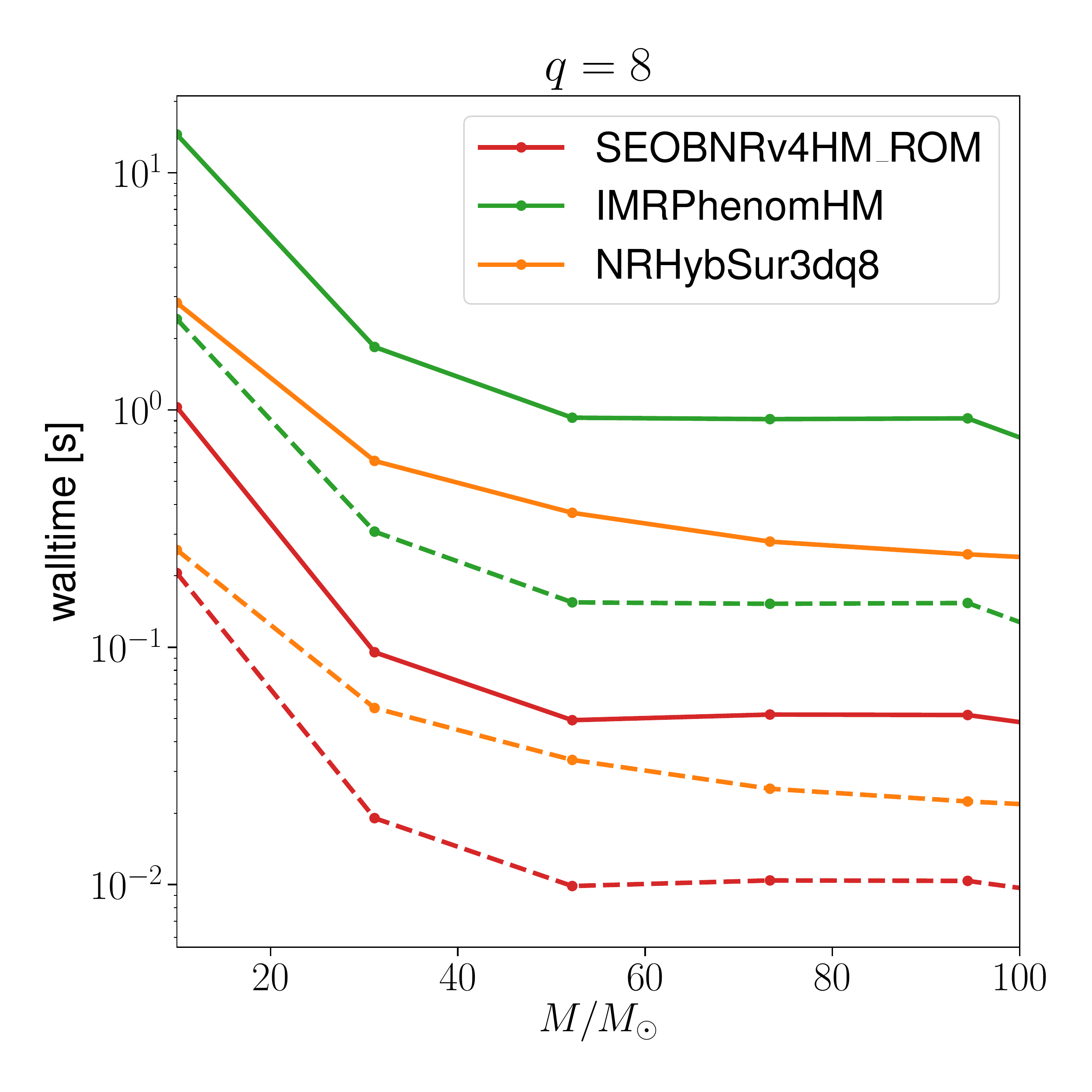}
   \caption{Walltime comparison between different spin-aligned waveform models with higher-order modes as a function of the total mass and for $q = 8$. The dashed lines indicate the walltime normalized by the number of modes included in the model respectively 5 fo \texttt{SEOBNRv4HM\_ROM}, 6 for \texttt{IMRPhenomHM} and 11 for \texttt{NRHybSur3dq8}.
   }
   \label{fig:walltimeq8}
\end{figure}

\subsubsection{Walltime comparison}
\label{sec:walltime}

We now perform a comparison of the walltime for generating a waveform between \texttt{SEOBNRv4HM\_ROM} and two waveform models that also include higher-order modes, namely \texttt{IMRPhenomHM}~\cite{London:2017bcn} and \texttt{NRHybSur3dq8}~\cite{Varma:2018mmi}.
As in Sec.~\ref{sec:speedup} we define walltime as the time to produce a frequency domain waveform at the same parameters. Since \texttt{NRHybSur3dq8} is a time domain model, we first generate the waveform in the time domain at a sample rate of 16384 Hz and then we compute its Fourier transform. For IMRPhenomHM and \texttt{SEOBNRv4HM\_ROM} the waveforms are generated in the frequency domain with a maximum frequency of 8192 Hz and a sampling interval of $1/T$ where $T$ is the duration in seconds of the associated time domain waveform.
The waveform models \texttt{SEOBNRv4HM\_ROM}, \texttt{IMRPhenomHM} and \texttt{NRHybSur3dq8} include a different numbers of modes in the waveform, respectively five $\left[(\ell, |m|) = (2,2),\allowbreak (2,1),\allowbreak (3,3),\allowbreak (4,4),\allowbreak (5,5)\right]$, six $[(\ell, |m|) = (2,2),\allowbreak (2,1),\allowbreak (3,3), \allowbreak (3,2), \allowbreak (4,4), \allowbreak (4,3)]$ and eleven $[(\ell,|m|) = (2,2), \allowbreak (2,1), \allowbreak (2,0), \allowbreak (3,3), \allowbreak (3,2), \allowbreak (3,1), \allowbreak (3,0), \allowbreak (4,4), \allowbreak (4,3),\allowbreak (4,2), \allowbreak (5,5)]$ modes. Since the total walltime is an increasing function of the number of modes, we also compute walltimes normalized by the number of modes to factor out this effect.
In Fig.~\ref{fig:walltimeq8} we show the walltime for generating a waveform with the different models as a function of the total mass for $q = 8$. \texttt{SEOBNRv4HM\_ROM} has walltimes of $\mathcal{O}(10)$ ms and is roughly 10 times faster than \texttt{IMRPhenomHM} or \texttt{NRHybSur3dq8}. When normalizing the walltime to the number of modes \texttt{SEOBNRv4HM\_ROM} is still about 10 times faster than \texttt{IMRPhenomHM}, but only $\sim 3$ times faster than \texttt{NRHybSur3dq8}.

\subsection{Parameter estimation study}
\label{sec:PEsec}

In this section we use the \texttt{SEOBNRv4HM\_ROM} model in a parameter estimation application.
For this purpose we create two mock signals (or injections) with the same binary parameters, using either \texttt{SEOBNRv4HM} or \texttt{NRHybSur3dq8} to generate the waveform. We then use \texttt{SEOBNRv4HM\_ROM}, \texttt{SEOBNRv4\_ROM}, and, as a comparison between waveform models for the second case, \texttt{IMRPhenomHM}~\cite{London:2017bcn}\footnote{A new version of the IMRPhenom waveform model with higher-order modes became only very recently available (see Refs.~\cite{Pratten:2020fqn,Garcia-Quiros:2020qpx}), therefore we have not been able to include it in our study. We defer comparisons with this model to future analysis.} and \texttt{NRHybSur3dq8} to compute posterior distributions from the mock signals. The analysis of the first mock signal will demonstrate the improvements in measuring binary parameters when using a model with higher harmonics with respect to a model including only the dominant $(\ell, |m|) = (2,2)$ mode.
The analysis of the second mock signal will give us a sense of possible biases due to modeling errors in the original \texttt{SEOBNRv4HM} model with respect to NR-surrogate  waveforms, which are close to NR simulations. In creating the mock signals we do not add detector noise. This choice is made to avoid additional uncertainty and bias introduced by a random noise realization. It is the natural choice given that the goal of this parameter estimation analysis is to check for possible biases due to inaccuracies in waveform models.

\subsubsection{Setup}

We choose parameters for the mock signals in order to emphasize the effect of higher-modes in the waveform.
Since the contribution of higher-order modes in the emitted \acp{GW} increases with the mass ratio, we use for the mock signals $q = 8$, the largest mass ratio available for the model \texttt{NRHybSur3dq8}. For the total mass we use $M = 67.5 M_\odot$ such that the values of the component masses $m_1 = 60 M_\odot$ and $m_2 = 7.5 M_\odot$ are consistent with the masses of BBH systems observed during O1 and O2 (see~\cite{LIGOScientific:2018mvr} and~\cite{LIGOScientific:2018jsj}). 
The waveform models are restricted to non-precessing spins and we pick $\chi_{1z} = 0.5$ and $\chi_{2z} = 0.3$. To maximize the effect of the higher-modes we inject the signal at edge-on inclination $(\iota = \pi/2)$ with respect to the observer. The coalescence phase $\phi_c$ is chosen to be 1.2 rad, while the polarization phase $\psi$ is set to 0.7 rad. The signal has been injected at gps-time 1249852257 s with a sky-position defined by its right ascension of 0.33 rad and its declination of -0.6 rad. Finally the distance of the mock signal is chosen by demanding a network-SNR of $21.8$ in the three detectors (LIGO Hanford, LIGO Livingston and Virgo) when using the Advanced LIGO and Advanced Virgo PSD at design sensitivity~\cite{Barsotti:2018}. The resulting distance is 627 Mpc.
We used \texttt{PyCBC}'s \texttt{pycbc\_generate\_hwinj}~\cite{alex_nitz_2020_3630601} to prepare the mock signal. To carry out Bayesian parameter estimation we used the Markov chain Monte Carlo code \texttt{LALInferenceMCMC}~\cite{Veitch:2014wba,LALInference}.
We choose a uniform prior in component masses in the range $[3,100] M_\odot$. Aligned component spins are assumed to be uniform in $[-1,1]$. The priors on the other parameters are the standard ones described in Appendix C.1 of Ref.~\cite{LIGOScientific:2018mvr}.

\subsubsection{Results}

\begin{figure*}[hbt]
  \centering
  \includegraphics[width=0.45\textwidth]{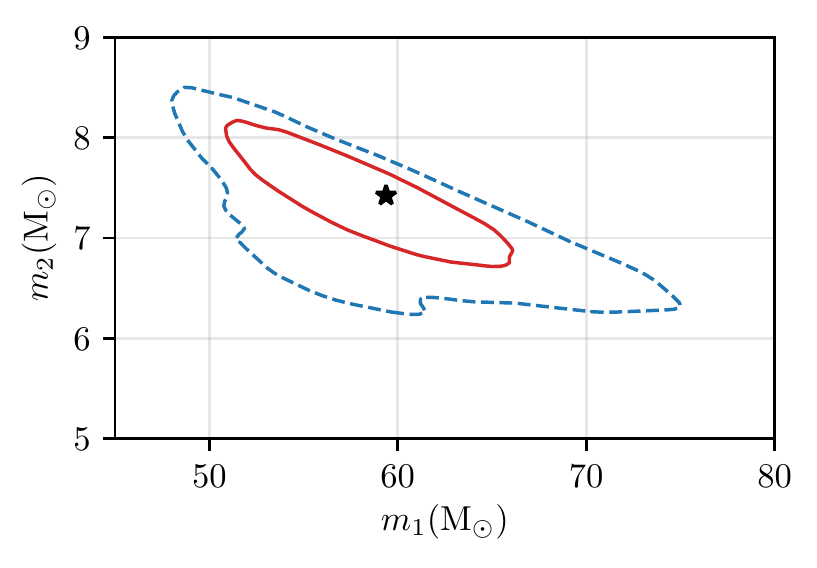}
  \includegraphics[width=0.45\textwidth]{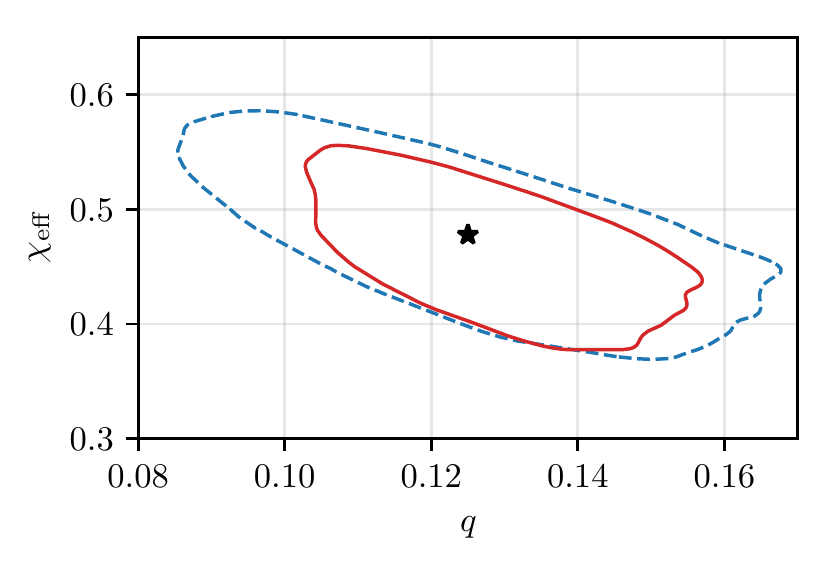}
  \includegraphics[width=0.45\textwidth]{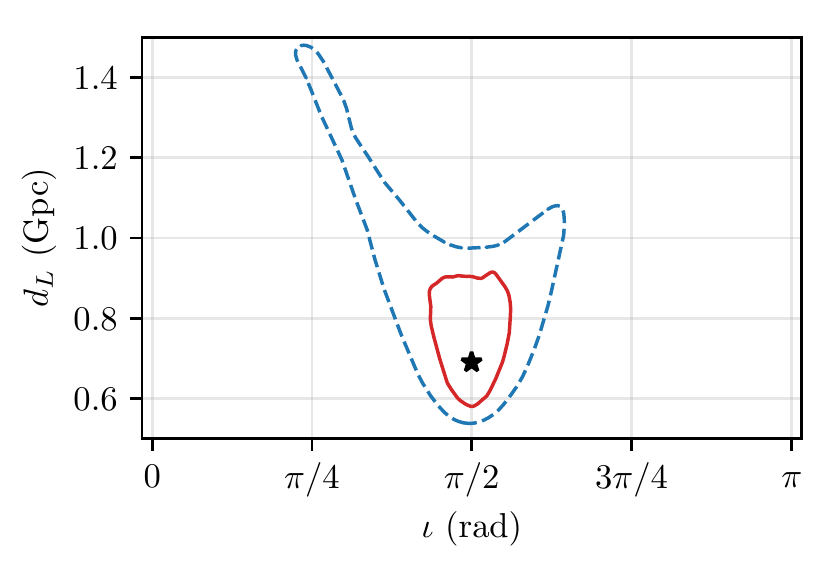}
  \includegraphics[width=0.45\textwidth]{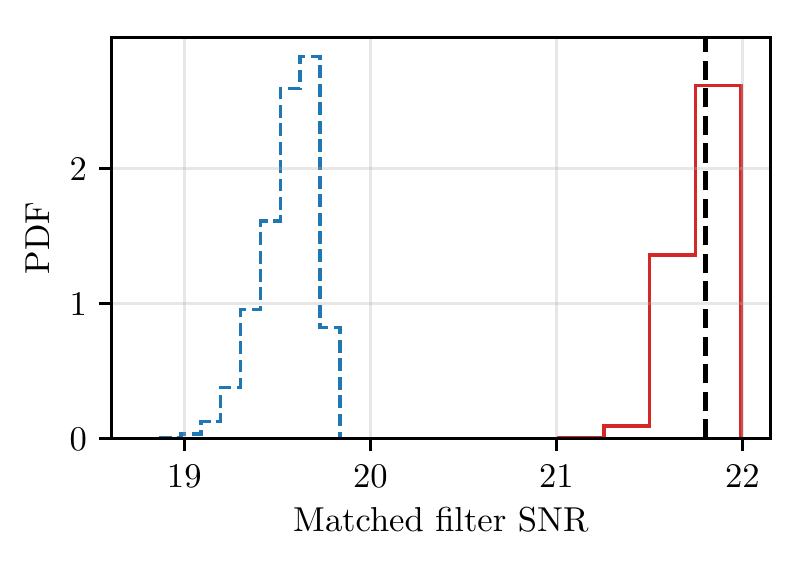}
  \includegraphics[trim={5.8cm 1cm 0 0}, clip, width=0.7\textwidth]{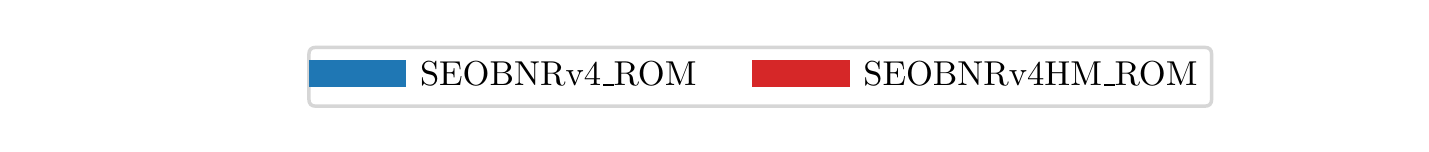}
	\caption{
  90\% credible regions and histograms of posterior distributions for a $q = 8$ \ac{BBH}. 
  The signal waveform is \texttt{SEOBNRv4HM\_ROM} and the stars represent binary parameters used for the signal. The mock signals are recovered with
  \texttt{SEOBNRv4\textunderscore ROM} and \texttt{SEOBNRv4HM\textunderscore ROM} waveform models.
 \emph{Top Left:} component masses in the source frame
  \emph{Top Right:} mass-ratio and effective aligned spin parameter.
  \emph{Bottom Left:} inclination angle and luminosity distance.
  \emph{Bottom Right:} matched filter \ac{SNR}
  }
  \label{fig:PE_all_EOBinj}
\end{figure*}

\begin{figure*}[hbt]
  \centering
  \includegraphics[width=0.45\textwidth]{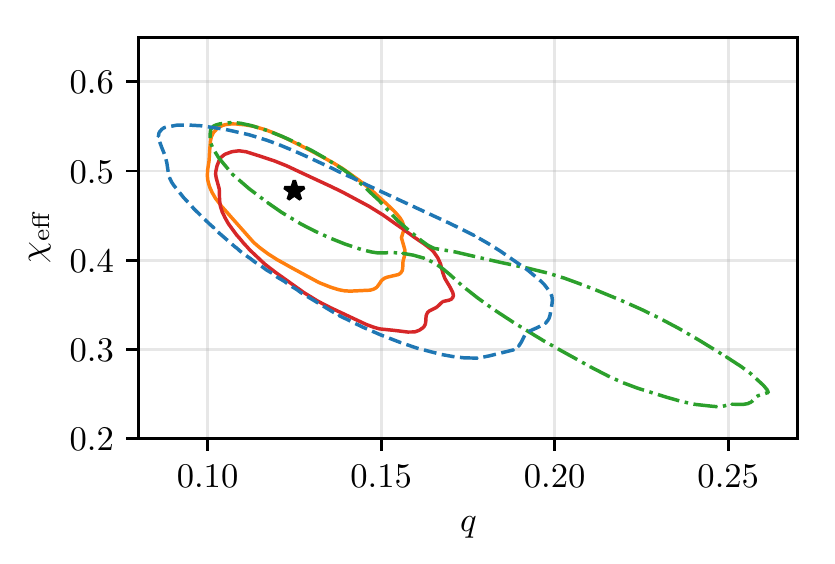}
  \includegraphics[width=0.45\textwidth]{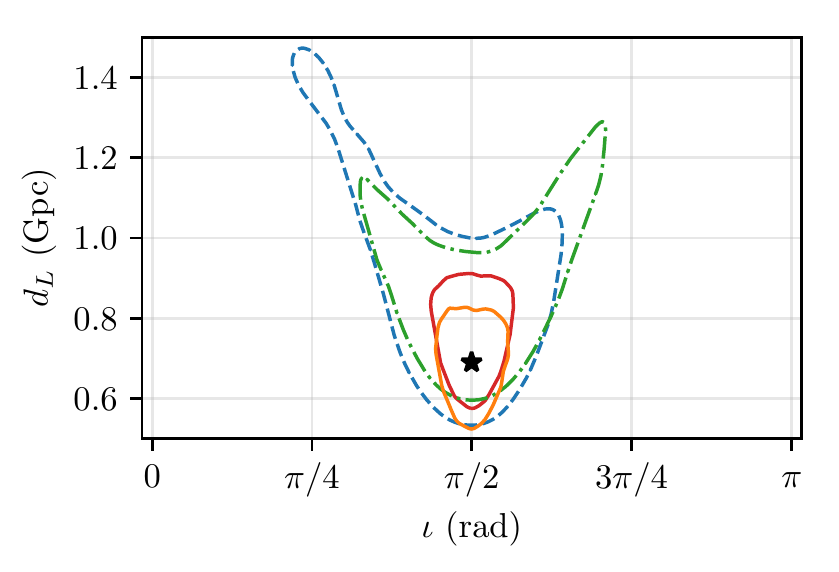}
  \includegraphics[trim={-1.0cm 1cm 2cm 0cm}, clip,width=0.9\textwidth]{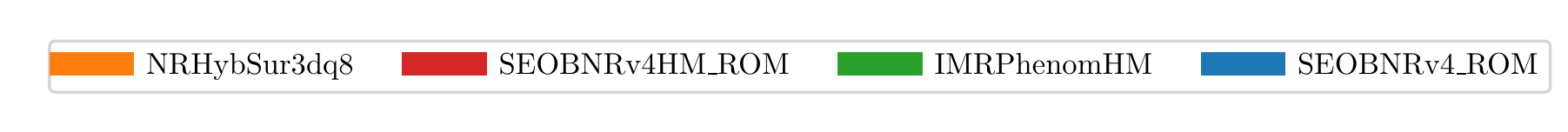}
	\caption{
  90\% credible regions and histograms of posterior distributions for a $q = 8$ \ac{BBH}. 
  The signal waveform is \texttt{NRHybSur3dq} and the stars represent binary parameters used for the signal. The mock signals are recovered with
  \texttt{SEOBNRv4\textunderscore ROM}, \texttt{SEOBNRv4HM\textunderscore ROM}, \texttt{IMRPhenomHM}, and  \texttt{NRHybSur3dq8} waveform models.
  \emph{Left:} mass-ratio and effective aligned spin parameter.
  \emph{Right:} inclination angle and luminosity distance.
  }
  \label{fig:PE_all_NRinj}
\end{figure*}

Let us first focus on the case in which the mock signal is generated with \texttt{SEOBNRv4HM}. In Fig.~\ref{fig:PE_all_EOBinj} we summarize the results of the parameter estimation analysis for some relevant binary parameters.
The top left panel shows the marginalized 2D posterior for the component source-frame masses, and the top right panel the marginalized 2D posterior for the mass ratio $q$ and the spin parameter $\chi_{\mathrm{eff}} = (m_1 \chi_1 + m_2 \chi_2) / (m_1 + m_2)$. In the bottom left panel we present the marginalized 2D posterior with inclination $\iota$ and luminosity distance $d_L$ and, finally, in the bottom right panel, we report the matched filter SNR. The star in the plots corresponds to the true value used for the mock signal, while the 2D contours of the posterior distributions represent $90\%$ credible regions. The waveform templates used to infer binary parameters are \texttt{SEOBNRv4\_ROM} (blue curve) and \texttt{SEOBNRv4HM\_ROM} (red curve). It is clear from the plots that all the  parameters reported in Fig.~\ref{fig:PE_all_EOBinj} are more precisely measured when using \texttt{SEOBNRv4HM\_ROM} instead of \texttt{SEOBNRv4\_ROM}. The posterior volume represents the degeneracy of the gravitational wave signal, and, in the absence of detector noise, this degeneracy is intrinsic to the waveforms. The inclusion of higher harmonics in \texttt{SEOBNRv4HM\_ROM} breaks the degeneracy between the parameters $q-\chi_{\mathrm{eff}}$ and $\iota-D_L$ and allows to measure them more precisely. These results are consistent with what was previously found in the literature~\cite{Graff:2015bba,Shaik:2019dym,Kalaghatgi:2019log}. As expected \texttt{SEOBNRv4HM\_ROM} also measures a larger matched filter SNR.

Let us now consider the case in which the mock signal is represented by \texttt{NRHybSur3dq8}. In Fig.~\ref{fig:PE_all_NRinj} we show the marginalized 2D posterior for the mass ratio $q$ and the spin parameter $\chi_{\mathrm{eff}}$ (left panel) and the marginalized 2D posterior with inclination $\iota$ and luminosity distance $d_L$ (right panel) as measured by the waveform models \texttt{SEOBNRv4HM\_ROM} (red curve), \texttt{SEOBNRv4\_ROM} (blue curve), \texttt{IMRPhenomHM} (green curve), and \texttt{NRHybSur3dq8} (orange curve). As before the star in the plots corresponds to the true value used for the mock signal, while the 2D contours of the posterior distributions represent $90\%$ credible regions. From the plots in Fig.~\ref{fig:PE_all_NRinj} it is clear that, as before, \texttt{SEOBNRv4HM\_ROM} recovers the binary parameters more precisely than \texttt{SEOBNRv4\_ROM}.
It is important to highlight that with \texttt{SEOBNRv4HM\_ROM} the binary parameters are recovered inside the $90\%$ credible regions. This means that for this quite asymmetric system at a moderately high SNR of $\sim 20$ the bias due to modeling errors in the original \texttt{SEOBNRv4HM} model compared to NR waveforms is negligible with respect to the statistical uncertainty.
In contrast, the marginal posterior distributions recovered for \texttt{IMRPhenomHM} are in general broader compared to the ones recovered by \texttt{SEOBNRv4HM\_ROM}, are notably bimodal in mass-ratio and effective spin, and extend a lot further along the line of $q$ - $\chi_\mathrm{eff}$ degeneracy. In distance and inclination the \texttt{IMRPhenomHM} posterior shows little improvement over \texttt{SEOBNRv4\_ROM} which does not include higher harmonics.
Finally, the marginal posteriors for \texttt{NRHybSur3dq8} are quite similar in size to those for \texttt{SEOBNRv4HM\_ROM}, but better centered around the true parameter values. This is as expected since the likelihood should peak at the true parameter values when the signal and template use the same waveform. The mock signal is also sufficiently loud for the posteriors to be likelihood- rather than prior-dominated, resulting in an unbiased parameter recovery.

This study shows that using \texttt{SEOBNRv4HM\_ROM} for parameter estimation yields unbiased measurements of the binary parameters at moderately high SNR even in a configuration where the effect of higher harmonics in the waveform is large. We defer a more comprehensive analysis to future studies.

\section{Conclusion}
\label{sec:conclusion}

In this paper we have presented a fast and accurate \ac{ROM} or surrogate model for the time domain \texttt{SEOBNRv4HM} \ac{EOB} waveform~\cite{Cotesta:2018fcv}. This model assumes spins aligned with the orbital angular momentum of the binary and includes the $(\ell, |m|) = (2, 1), (3, 3), (4, 4), (5, 5)$ spherical harmonic modes beyond the dominant $(\ell, |m|) = (2, 2)$ mode.

While the construction of this Fourier domain \ac{ROM} broadly follows previous work~\cite{Purrer:2014fza,Purrer:2015tud} we have introduced the following new features to accurately represent the higher harmonics and make the model more flexible (see Sec.~\ref{sec:techniques}).
While previous models used an amplitude / phase decomposition of the Fourier domain waveform, we here define a carrier signal (see Eq.~\eqref{eq:defcarrier}) based on the time domain orbital phase. Subsequently we extract the carrier phasing from each Fourier domain waveform mode (see Eq.~\eqref{eq:defhcoorb}). This essentially makes the phase of modes almost constant in the inspiral, defining what we here call ``coorbital modes''. This choice allows us to avoid zero-crossings in the subdominant harmonics which could spoil the smoothness of the training data and make accurate interpolation of the waveform data over parameter space very difficult.
We perform alignment in the time domain to keep track of the time of coalescence of the training set waveforms and this information is preserved in the \ac{ROM}.
We use here an alternative approach to dealing with the fact that the ringdown frequency varies over the parameter space, but waveform data needs to be given on a common frequency grid to build a \ac{ROM}. We rescale the geometric frequency parameter so that the ringdown is reached before a fixed termination frequency which demarcates the end of the frequency grid. We use the inverse rescaling during the evaluation of the \ac{ROM}.
We extend the \ac{ROM} to arbitrarily low frequencies by splicing it together with multipolar \ac{PN} waveforms. Therefore, it can in principle be used for arbitrarily light compact binary systems.
We decompose waveform input data in orthonormal bases using the \ac{SVD}, and build a model by constructing a tensor product spline over the 3-dimensional parameter space of mass-ratio and the two aligned spins of a binary. To increase model accuracy and efficiency we use domain decomposition in frequency and in parameter space.

In Sec.~\ref{sec:accuracy} we demonstrate that the \ac{ROM} has a very high faithfulness (or match) with \texttt{SEOBNRv4HM}. Maximizing over inclination, reference phase and effective polarization of the source waveform (see Eq.~\eqref{eq:max_unfaith}) the maximum mismatch over the remaining source parameters is below $0.03\%$ for binaries with a total mass below $100 M_\odot$ and below $0.2\%$ for binaries at $300 M_\odot$ (see Table~\ref{tab:maximum_unfaith_M_summary_table}). Even for this very conservative choice the mismatch is at least an order of magnitude lower than the unfaithfulness of \texttt{SEOBNRv4HM} against \ac{NR} simulations. Therefore the additional modeling error introduced in building the \ac{ROM} is strongly subdominant and the \ac{ROM} very accurately represents the \texttt{SEOBNRv4HM} waveform model.
In Sec.~\ref{sec:computational_performance} we show that our \ac{ROM} accelerates waveform evaluation by a factor 100 -- 200 compared to \texttt{SEOBNRv4HM} and favorably compares against other higher mode waveform models for \ac{BBH} systems, being about an order of magnitude faster.
We showcase in Sec.~\ref{sec:PEsec} (see Figs.~\ref{fig:PE_all_EOBinj}, and~\ref{fig:PE_all_NRinj}) that our \ac{ROM} can recover component masses and spins, and especially distance and inclination angle for a quite asymmetric and spinning \ac{BBH} with increased precision compared to the \texttt{SEOBNRv4\_ROM} waveform which only models the dominant mode. In addition we show that the \ac{ROM} accurately recovers binary parameters, irrespective of whether the source is represented by a \texttt{SEOBNRv4HM} or a \texttt{NRHybSur3dq8} waveform.
Our \ac{ROM} gives a significantly more accurate parameter recovery compared to the phenomenological \texttt{IMRPhenomHM} waveform and is close to the \texttt{NRHybSur3dq} NR-surrogate model, while being more versatile and covering a significantly larger parameter space.

This \ac{ROM} should prove a very useful tool for \ac{GW} data analysis to describe systems
where the contribution of higher harmonics is important in terms of additional signal-to-noise-ratio and discriminating power for detection and parameter inference.
We stress that the ROM is very fast and reproduces the \texttt{SEOBNRv4HM} model with a great accuracy over the widest range in parameter space of all inspiral-merger-ringdown higher mode models available to date, from mass-ratio 1 to 1:50 where aligned spins can take values in the full range allowed for Kerr \acp{BH}, up to extremal spins.

\begin{acknowledgments}
The authors would like to thank Stas Babak, Alessandra Buonanno, Nils Fischer, Bhooshan Gadre, Jonathan Gair, Cecilio Garcia-Quiros, Steffen Grunewald, Ian Harry, Serguei Ossokine, Harald Pfeiffer and Vijay Varma for useful discussions.
The authors acknowledge usage of AEI's Hypatia computer cluster.

This is LIGO Document Number LIGO-P2000106.

\end{acknowledgments}



\bibliography{paper}
\end{document}